%% file: bch-v4.tex
\documentclass[11pt]{article}
\usepackage{epsf}
\usepackage{color}
\usepackage{epsfig}

\usepackage{graphicx}
\usepackage{amsmath}
\usepackage{latexsym}
\usepackage{amssymb}

\usepackage{a4}

\newcommand{\ffrac}[2]{#1/#2}

\newtheorem{theorem}{Theorem}[section]

\newtheorem{algorithm}{Algorithm}

\numberwithin{equation}{section}

\newcommand{\bracketab}{[\alpha,\beta]}
\newcommand{\ad}{{\rm ad}}

\newcommand{\M}{\ensuremath{\mathcal{M}}}
\newcommand{\XM}{\ensuremath{\mathfrak{X}(\M)}}

\newcommand{\Z}{\mathbb{Z}^+}
\newcommand{\R}{\mathbb{R}}
\newcommand{\Q}{\mathbb{Q}}
\newcommand{\g}{\mathfrak{g}}

\renewcommand{\L}{\mathcal{L}}
\newcommand{\cir}{\, {\scriptstyle \circ}\, }
\newcommand{\T}{\mathcal{T}}
\newcommand{\e}{\ensuremath{\mathrm{e}}}

\usepackage{pspicture}
\input{2trees}

\newlength{\treewidth}
\newlength{\treeheight}

\newcommand{\tb}[2][-0.3]{
        \setlength{\unitlength}{1.3ex}
        \settowidth{\treewidth}{#2}%
        \settoheight{\treeheight}{#2}%
        \raisebox{#1%
        \treeheight}{\mbox{\hspace{0.5\unitlength}#2[1.5ex]\hspace{0.5\unitlength}}}}

\begin{document}

\title{An efficient  algorithm for computing the Baker--Campbell--Haus\-dorff
series and some of its applications}

\author{Fernando Casas$^{1}$\thanks{Corresponding author. Email: \texttt{Fernando.Casas@uji.es}}
   \and
Ander Murua$^{2}$\thanks{Email: \texttt{Ander.Murua@ehu.es}}
   }

\date{}
\maketitle

\begin{abstract}

We provide a new algorithm for generating the Baker--Campbell--Haus\-dorff (BCH) series
$Z = \log(\e^X \e^Y)$ in an arbitrary generalized Hall basis of the free Lie algebra
$\mathcal{L}(X,Y)$ generated by $X$ and $Y$. It is based on the close relationship
of $\mathcal{L}(X,Y)$ with a Lie algebraic structure of labeled rooted trees. With this algorithm,
the computation of the BCH series up to degree 20 (111013 independent elements in  
$\mathcal{L}(X,Y)$) takes less than 15 minutes on a personal
computer and requires 1.5 GBytes of memory. We also address the issue of
the convergence of the series,
providing an optimal convergence domain when $X$ and $Y$ are real or complex
matrices.

\vspace*{0.5cm}

\begin{description}
 \item $^1$Departament de Matem\`atiques, Universitat Jaume I,
  E-12071 Castell\'on, Spain.
 \item $^2$Konputazio Zientziak eta A.A. saila, Informatika
Fakultatea, EHU/UPV, Donostia/San Sebasti\'an, Spain.
\end{description}

\end{abstract}

\section{Introduction}   \label{sec.1}

The Baker--Campbell--Hausdorff formula deals with the expansion of $Z$ in
$\e^X \, \e^Y = \e^Z$ in terms of nested commutators of $X$ and $Y$ when they
are assumed to be non-commuting operators. If we  introduce the
formal series for the exponential function
\begin{equation}   \label{neq.1}
   \e^X \, \e^Y = \sum_{p,q = 0}^{\infty} \frac{1}{p! \, q!}
            X^p \, Y^q
\end{equation}
and substitute this series in the formal series defining the
logarithm function
\[
   \log Z = \sum_{k=1}^{\infty} \frac{(-1)^{k-1}}{k} (Z-1)^k
\]
one obtains
\[
  \log( \e^X \, \e^Y) = \sum_{k=1}^{\infty} \frac{(-1)^{k-1}}{k}
   \sum \frac{X^{p_1} Y^{q_1} \ldots X^{p_k} Y^{q_k}}{p_1! \,
      q_1! \ldots p_k! \, q_k!},
\]
where the inner summation extends over all non-negative integers
$p_1$, $q_1$, \ldots, $p_k$, $q_k$ for which $p_i + q_i > 0$
($i=1,2, \ldots, k$). Gathering together the terms for which
$p_1+q_1+p_2+q_2+ \cdots +p_k + q_k = m$ we can write
\begin{equation}   \label{eq.1.1}
  Z = \log( \e^X \, \e^Y) = \sum_{m=1}^{\infty} P_m(X,Y),
\end{equation}
where $P_m(X,Y)$ is a homogeneous polynomial of degree $m$ in
the non-commuting variables $X$ and $Y$. Campbell \cite{campbell98oal}, 
Baker \cite{baker05aac} and Hausdorff \cite{hausdorff06dse}
addressed the question whether $Z$ can be represented as a series
of nested commutators of $X$ and $Y$, without producing a general formula. 
We recall here that the commutator $[X,Y]$ is defined as $XY - YX$.
It was
Dynkin \cite{dynkin47eot} who finally derived an explicit formula for $Z$ as
\begin{equation}  \label{eq.1.2}
  Z = \sum_{k=1}^{\infty} \sum_{p_i, q_i}
    \frac{(-1)^{k-1}}{k} \frac{[X^{p_1} Y^{q_1} \ldots
      X^{p_k} Y^{q_k} ]}{(\sum_{i=1}^k (p_i + q_i)) \,
      p_1! \, q_1! \ldots p_k! \, q_k!}.
\end{equation} 
Here the inner summation is taken over all non-negative
integers $p_1, q_1, \ldots$, $p_k$, $q_k$ such that
$p_1+q_1>0, \ldots, p_k + q_k > 0$ and 
$[ X^{p_1} Y^{q_1} \ldots X^{p_k} Y^{q_k}]$ denotes the right
nested commutator based on the \emph{word}
$ X^{p_1} Y^{q_1} \ldots X^{p_k} Y^{q_k}$.
Expression (\ref{eq.1.2}) is known, for obvious reasons, as  the {\it Baker--Campbell--Hausdorff 
series in the Dynkin form}.  By rearranging terms, it is clear that $Z$ can be written as 
\begin{equation}  \label{eq.1.3}
   Z = \log( \e^X \, \e^Y) = X + Y +
    \sum_{m=2}^{\infty} \, Z_m,
\end{equation}
with $Z_m(X,Y)$ a homogeneous Lie polynomial in $X$ and $Y$
of degree $m$, i.e., it is a $\mathbb{Q}$-linear combination
of commutators of the form $[V_1,[V_2, \ldots,[V_{m-1},V_m]
\ldots]]$ with $V_i \in \{X,Y\}$ for $1 \le i \le m$. The first terms read explicitly
 \begin{eqnarray*} 
   Z_2 & = & \frac{1}{2} [X,Y]   \\
   Z_3 & = & \frac{1}{12} [X,[X,Y]] - \frac{1}{12}
                 [Y,[X,Y]]    \\
   Z_4 & = & \frac{1}{24} [X,[Y,[Y,X]]]  
\end{eqnarray*}
The expression $\e^X \, \e^Y = \e^Z$ is then called  the \emph{Baker--Campbell--Hausdorff
formula}  (BCH for short), although other different labels (e.g., Campbell--Baker--Hausdorff, 
Baker--Hausdorff, Campbell--Hausdorff)  are commonly attached to it in the
literature. The formula (\ref{eq.1.2}) is certainly awkward to use due to the
complexity of the sums involved. Notice, in particular, that different choices of
$p_i$, $q_i$, $k$ in (\ref{eq.1.2}) may lead to terms in the same commutator. Thus, for instance,
$[X^3 Y^1] = [X^1 Y^0 X^2 Y^1] = [X,[X,[X,Y]]]$. An additional difficulty arises from the fact that
not all the commutators are independent, due to the Jacobi identity \cite{varadarajan84lgl}:
\[
    [X_1,[X_2,X_3]] + [X_2,[X_3,X_1]] + [X_3,[X_1,X_2]] = 0.  
\]    

The BCH formula plays a fundamental role in many fields of mathematics
(theory of linear differential
equations \cite{magnus54ote}, Lie groups  \cite{gorbatsevich97fol},
numerical analysis \cite{hairer06gni}), theoretical physics
(perturbation theory \cite{dragt76lsa}, Quantum Mechanics \cite{weiss62tbf},
Statistical Mechanics \cite{kumar65oet,wilcox67eoa}, quantum computing 
\cite{sornborger99hom}) and control theory (analysis and design of nonlinear
control laws, nonlinear filters, stabilization of rigid bodies \cite{torres05asp}). 
In particular, in the theory of Lie groups, with this formula 
one can explicitly write the operation of
multiplication in a Lie group in canonical coordinates in terms
of the Lie bracket operation in its tangent algebra and also
prove the existence of a local Lie group with a given Lie
algebra \cite{gorbatsevich97fol}.

Also in the numerical treatment of differential equations on
manifolds \cite{iserles00lgm,hairer06gni}, the BCH formula is quite useful.  
If $\M$ is a smooth manifold and $\XM$ denotes the linear
space of smooth vector fields on $\M$, then
a Lie algebra structure is established in $\XM$
by using the Lie bracket $[X,Y]$ of fields $X$ and $Y \in \XM$ \cite{varadarajan84lgl}. 
The flow of a vector field $X \in \XM$ is a mapping
$\exp(X)$ defined through the solution of the differential
equation
\begin{equation}   \label{de.1.1}
   \frac{du}{dt} = X(u),  \qquad\qquad u(0) = q \in \M
\end{equation}
as $\exp(t X)(q) = u(t)$. Many numerical methods used to
approximately solving equation (\ref{de.1.1})  are based on
compositions of maps that are flows of vector fields \cite{hairer06gni}. 
To be more specific, suppose the vector field $X$ can be split as
$X = A + B$ and that the flows corresponding to $A(u)$ and $B(u)$ can
be explicitly obtained. Then one may consider an approximation of the form
$\Psi_h \equiv \exp(h a_1 A) \exp(h b_1 B) \cdots \exp(h a_k A) \exp(h b_k B)$ 
for the exact flow $\exp(h(A+B))$ of (\ref{de.1.1}) after a time step $h$. The idea
now is to obtain the conditions to be satisfied by the coefficients $a_i$, $b_i$ so that
$\Psi_h(q) = u(h) + \mathcal{O}(h^{p+1})$ as $h \rightarrow 0$, and this can be done by applying the BCH
formula in sequence to the expression of $\Psi$ up to the degree required
by the order of approximation $p$ \cite{mclachlan02sm}. This task can be carried out
quite easily provided one has explicit expressions of $Z_m$ implemented in a symbolic 
algebra package \cite{koseleff93cfp,torres05asp}.

In addition to the Dynkin form (\ref{eq.1.2}), there are other standard 
procedures to construct explicitly the BCH series.  Recall that the free Lie algebra  $\L(X,Y)$ generated by the symbols $X$ and $Y$ can be considered as a subspace (the subspace of Lie polynomials) of the vector space spanned by the words $w$ in the symbols $X$ and $Y$, i.e., 
$w = a_1 a_2 \ldots a_m$, each $a_i$ being $X$ or $Y$.
Thus, the BCH series admits
the explicit associative presentation 
     \begin{equation}   \label{eq.2.1}
  Z = X + Y + \sum_{m=2}^{\infty} \, \sum_{w, |w|=m} \,
     g_w \, w,
\end{equation}
in which $g_w$
is a rational coefficient and the inner sum is taken over all words
$w$ with length $|w|=m$. Here the length of $w$ is the number
of letters it contains. The coefficients can be computed with a
procedure based on a family
of recursively computable polynomials \cite{goldberg56tfp}.

Although the terms in equation (\ref{eq.2.1}) are expressed as linear combinations of individual words (which are not Lie polynomials), by virtue of the Dynkin--Specht--Wever
theorem \cite{jacobson79lal}, $Z$ can be written as 
\begin{equation}   \label{eq.2.2}
  Z = X + Y + \sum_{m=2}^{\infty} \, \frac{1}{m} \, \sum_{w, |w|=m} \,
     g_w \, [w],
\end{equation}
that is, the individual terms are the same as in the associative
series (\ref{eq.2.1}) except that the word $w= a_1 a_2 \ldots a_m$
is replaced with the right nested
commutator $[w] = [a_1,[a_2, \ldots [a_{m-1},a_m]\ldots]]$ and
the coefficient $g_w$ is divided by the word length $m$ \cite{thompson82cra}.
This gives explicit expressions of the terms $Z_m$ in the BCH series (\ref{eq.1.3})  as a linear combination of nested commutators of homogeneous degree, that is, as a linear combination of elements of the homogeneous subspace $\L(X,Y)_m$ of degre $m$ of the free Lie algebra  $\L(X,Y)$. However, it should be stressed that the set of nested commutators $[w]$ 
for words $w$ of length $m$ is \emph{not} a basis of the homogeneous subspace $\L(X,Y)_m$.

By introducing a parameter $\tau$ and differentiating with respect to $\tau$
the power series
$\sum_{m \ge 1} \tau^m Z_m = \log(\exp(\tau X) \exp(\tau Y))$,
the following recursion formula is derived in \cite{varadarajan84lgl}:
\begin{eqnarray}   \label{varada}
      Z_1 & = & X + Y  \\
%      (m+1) Z_{m+1} & = & \frac{1}{2} [X-Y, Z_m]  + \sum_{p=1}^{[m/2]} \frac{B_{2p}}{(2p)!} 
%       \sum [Z_{k_1}, [\cdots [Z_{k_{2p}}, X+Y] \cdots ]],
%      \quad m \ge 1  \nonumber
      m Z_{m} & = & \frac{1}{2} [X-Y, Z_{m-1}]  + \sum_{p=1}^{[(m-1)/2]} \frac{B_{2p}}{(2p)!} \left(\ad_{Z}^{2p} (X+Y)\right)_{m},
      \qquad m \ge 1.  \nonumber
\end{eqnarray}
Here $Z= \sum_{m \ge 1} Z_m$, $\mathrm{ad}_{Z}^k (X+Y) = [Z, \mathrm{ad}_{Z}^{k-1} (X+Y)]$, 
the $B_{j}$ stand for the Bernoulli numbers \cite{abramowitz65hom}, 
and $\left(\mathrm{ad}_{Z}^{2p} (X+Y)\right)_{m}$ denotes the projection 
of $\ad_{Z}^{2p} (X+Y)$ onto the homogeneous subspace $\L(X,Y)_m$, 
which can be written in terms of $Z_1$, $Z_2$, $Z_3$, $\ldots$ as
\begin{eqnarray*}
  \left(\ad_{Z}^{2p} (X+Y)\right)_{m} =
    \sum_{
            k_1 + \cdots + k_{2p} = m-1 \atop
            k_1 \ge 1, \ldots, k_{2p} \ge 1}
  [Z_{k_1}, [\cdots [Z_{k_{2p}}, X+Y] \cdots ]].
\end{eqnarray*}

Explicit formulas (\ref{eq.1.2}), (\ref{eq.2.2})  as well as recursion (\ref{varada}) 
can be used in principle to construct the BCH series up to arbitrary degree in terms of
commutators.  As a matter of fact, several systematic computations of the series have been
carried out along the years, starting with the work of Richtmyer and Greenspan in 1965 \cite{richtmyer65eot},  where 
results up to degree eight are reported. Later on, Newman and
Thompson obtained the coefficients $g_w$ in (\ref{eq.2.2}) up to words of length 20
\cite{newman87nvo}, Bose \cite{bose89dmo}
constructed an algorithm to compute directly the coefficient
of a given commutator in the Dynkin presentation (\ref{eq.1.2}) and Oteo \cite{oteo91tbc} and
Kolsrud \cite{kolsrud93mri}
presented a
simplified expression of (\ref{eq.1.2}) in terms of right nested commutators up to degree eight and
nine, respectively. More recently, Reinsch \cite{reinsch00ase} 
has proposed a matrix operation procedure for
calculating the polynomials $P_m(X,Y)$ in (\ref{eq.1.1}) which can be easily implemented in
any symbolic algebra package. Again, the Dynkin--Specht--Wever has to be used to write
the resulting expressions in terms of commutators.

As mentioned before, all of these procedures exhibit a key limitation, 
however: the iterated commutators are not 
all linearly independent due to the Jacobi identity (and other identities involving 
nested commutators of higher degree which are originated by it \cite{oteo91tbc}). 
In other words, they do not provide
expressions directly in terms of a basis of the free Lie algebra $\mathcal{L}(X,Y)$. This is required, for instance, in applications of the BCH formula in the numerical integration of ordinary differential equations, or when one wants to study specific features of the series, such as the distribution of the coefficients and other combinatorial properties \cite{newman87nvo}.

Of course, it is always possible to
express the resulting formulas in terms of a basis of $\mathcal{L}(X,Y)$ 
but this rewriting process is very time consuming and requires a good deal of memory
resources. In practice, going beyond degree $m=11$ constitutes a difficult task indeed
\cite{michel74bda,koseleff93cfp,torres05asp}, since the number of terms involved in the series grows exponentially. 

Our goal is then to express the BCH series as
\begin{eqnarray}
  \label{eq:BCHEi}
 Z= \log(\exp(X)\exp(Y)) = \sum_{i\geq 1} z_i\, E_i,
\end{eqnarray}
where $z_i \in \Q$ ($i\geq 1$) and $\{E_i \ : \ i=1,2,3,\ldots \}$ is a basis of $\L(X,Y)$ 
whose elements are of the form
\begin{eqnarray}
  \label{eq:Ei}
E_1=X, \quad E_2=Y, \quad \mbox{and} \quad   E_i = [E_{i'},E_{i''}] \quad  i\geq 3,
\end{eqnarray}
for appropriate values of the integers $i', i''<i$ (for $i=3,4,\ldots$). Clearly, each $E_i$ in (\ref{eq:Ei}) is a homogeneous Lie polynomial of  degree $|i|$, where 
\begin{eqnarray}
\label{eq:degree}
|1|=|2|=1, \quad \mbox{and} \quad |i|=|i'|+|i''| \quad  \mbox{for} \quad i \geq 3.
\end{eqnarray}
 We will focus on a general class of bases of the free Lie algebra $\L(X,Y)$, referred to
 in the current literature as generalized Hall bases and also as
 Hall--Viennot bases~\cite{reutenauer93fla,viennot78adl}. These include the Lyndon basis~\cite{lothaire83cow,viennot78adl},  and different variants of the classical Hall basis (see \cite{reutenauer93fla} for references). 
Specifically, in this paper we present a new procedure to write
the BCH series (\ref{eq:BCHEi}) for
an arbitrary Hall--Viennot basis. Such an algorithm is based on results
obtained in \cite{murua06tha}, in particular those relating a certain
Lie algebra structure $\g$ on rooted trees with the description of
a free Lie algebra in terms of a Hall basis. This Lie algebra $\g$ on
rooted trees was first considered in \cite{dur86mfi}, whereas a closely related
Lie algebra on labeled rooted trees was treated in \cite{grossman89has}
(see \cite{hoffman03cof} for the relation of these two Lie algebras and
for further references about related algebraic structures on rooted trees).

We have implemented the algorithm in \textit{Mathematica} (it can
also be programmed in Fortran
or C for more efficiency). The resulting procedure gives the BCH series up to a
prescribed degree directly in terms of a Hall--Viennot basis of
$\mathcal{L}(X,Y)$. As an illustration, obtaining the series (in the classical basis of P. Hall) up to degree
$m=20$ with a personal computer
(2.4 GHz Intel Core 2 Duo processor  with 2 GBytes of RAM) requires less than 15 minutes of CPU time and 1.5 GBytes of memory. The resulting expression has 109697 non-vanishing coefficients out of 111013 elements $E_i$ of degree $|i|\leq 20$ in the Hall basis. As far as we know, there
are no results up to such a high degree reported in the literature. For comparison with other
procedures, the authors of \cite{torres05asp} report 25 hours of CPU time and 
17.5 MBytes with a Pentium III  PC to achieve degree 10. By contrast, our
algorithm is able to achieve $m=10$ in $0.058$ seconds and only needs 5.4 MBytes of 
computer memory.

In Table~\ref{tab:BCH} in the Appendix, we give the values of $i'$ and $i''$ for the elements $E_i$ of degree $|i|\leq 9$ in the Hall basis and their coefficients $z_i$ in the BCH formula (\ref{eq:BCHEi}). The elements of the basis are ordered in such a way that $i<j$ if $|i|<|j|$, and the horizontal lines in the table separate elements of different homogeneous degree. Extension of Table~\ref{tab:BCH} up to terms of degree 20 is available at
the website \texttt{www.gicas.uji.es/research/bch.html}  
for both the basis of P. Hall and the Lyndon basis.
As an example, the last element of degree 20 in the Hall basis  is
\begin{eqnarray*}
  E_{111013} &=& [ [[[[Y,X],Y],[Y,X]],[[[Y,X],X],[Y,X]]], \\
           & & \phantom{[}[[[[Y,X],Y],[Y,X]],[[[[Y,X],Y],Y],Y]]],
\end{eqnarray*}
and the corresponding coefficient in (\ref{eq:BCHEi}) reads
\begin{eqnarray*}
  z_{111013} = -\frac{19234697}{140792940288}.
\end{eqnarray*}

Another central issue addressed in this paper concerns the convergence 
properties of the BCH series. Suppose
we introduce a sub-multiplicative norm $\| \cdot \|$ such that 
\begin{equation}   \label{eq.intro.1}
     \|[X,Y]\| \le \mu \|X\| \, \|Y\|
\end{equation}
for some $\mu > 0$. Then it is not difficult to show that the series (\ref{eq.1.2}) is
absolutely convergent as long as $\|X\| + \|Y\| < (\log 2) /\mu$ \cite{bourbaki89lga,suzuki77otc}. 
As a matter of fact, several improved bounds have been obtained for the different presentations.
Thus, in particular, the Lie presentation (\ref{eq.2.2}) 
converges absolutely if $\|X\| \le 1/\mu$ and $\|Y\| \le 1/\mu$ in a normed Lie algebra
$\mathfrak{g}$ with a norm satisfying (\ref{eq.intro.1}) \cite{newman89cdf,thompson89cpf}, 
whereas in \cite{blanes04otc} it has been shown that the series $Z = \sum_{m \ge 1} Z_m$
is absolutely convergent for all $X,Y$ such that 
\begin{equation}  \label{con.do.1}
        \mu \|X\| < \int_{\mu \|Y\|}^{2\pi} \frac{1}{2 + \frac{t}{2} (1 - \cot (\frac{1}{2} t))} \, dt
\end{equation}
and the corresponding expression obtained by interchanging in (\ref{con.do.1}) $X$ by $Y$. 
Moreover, the series diverges in general if $\|X\| + \|Y\| \ge \pi$ when $\mu = 2$ 
\cite{michel74bda}. Here we provide a generalization of this feature based on the well
known Magnus expansion for linear differential equations \cite{magnus54ote}
and also we give a more precise characterization of
the convergence domain of the series when $X$ and $Y$ are (real or complex) matrices.

\section{An algorithm for computing the BCH series based on rooted trees}
\label{section2}

\subsection{Summary of the procedure}

Our starting point is the vector space $\g$ of maps $\alpha:\T \rightarrow \R$, where
$\T$ denotes the set of rooted trees with black and
  white vertices 
\begin{eqnarray*}
  \T = \left\{ \tb{\nta}, \tb{\ntb}, \tb{\ntc}, \tb{\ntd},
  \tb{\nte}, \tb{\ntf}, \tb{\ntg}, \tb{\nth}, \tb{\nti}, \tb{\ntao},
  \ldots, \tb{\ntae}, \tb{\ntaf}, \tb{\ntag}, \tb{\ntah}, \ldots
  \right\}.
\end{eqnarray*}
In the combinatorial literature, $\T$ is typically referred to as the set
of labeled rooted trees with two labels, `black' and `white'.
Hereafter, we refer to the elements of $\T$ as bicoloured rooted trees.

The vector space $\g$ 
% of maps $\alpha:\T \to \R$ 
is endowed with a Lie
algebra structure by defining the Lie bracket $[\alpha,\beta] \in \g$ of 
two arbitrary maps $\alpha,\beta \in \g$ as follows. For each $u \in \T$,
\begin{eqnarray}
  \label{eq:bracket}
[\alpha,\beta](u) = \sum_{j=1}^{|u|-1}  \big(\alpha(u_{(j)}) \beta(u^{(j)}) - \alpha(u^{(j)}) \beta(u_{(j)}) \big),
\end{eqnarray}
where $|u|$ denotes the number vertices of $u$, and each of the  pairs of trees $(u_{(j)},u^{(j)}) \in \T \times \T$, $j=1,\ldots,|u|-1$, is obtained from $u$ by removing one of the $|u|-1$ edges of the rooted tree $u$, the root of $u_{(j)}$ being the original root of $u$.  For instance,
\begin{eqnarray}
  [\alpha,\beta](\tb{\nte}) &=& 
\alpha(\tb{\ntb}) \beta(\tb{\nta}) - \alpha(\tb{\nta}) \beta(\tb{\ntb}), 
\qquad\quad    [\alpha,\beta](\tb{\ntf}) = 0, 
\nonumber \\
    \bracketab(\tb{\ntah}) &=& 2 \big( \alpha(\tb{\nte}) \beta(\tb{\nta}) 
- \alpha(\tb{\nta}) \beta(\tb{\nte}) \big), \label{eq:[a,b]ex} \\
\nonumber
    \bracketab(\tb{\ntai}) &=&  
\alpha(\tb{\ntf}) \beta(\tb{\nta}) + \alpha(\tb{\nte}) \beta(\tb{\ntb})
-\alpha(\tb{\nta}) \beta(\tb{\ntf})
- \alpha(\tb{\ntb}) \beta(\tb{\nte}).
\end{eqnarray}

An important feature of the Lie algebra $\g$ is that
the Lie subalgebra of $\g$ generated by 
the maps $X,Y \in \g$ defined as
\begin{eqnarray}
  \label{eq:XY}
X(u)= \left\{
\begin{array}{lcl}
  1 & \mbox{if} & u=\tb{\nta} \\
   0 & \mbox{if} & u \in \T\backslash\{\tb{\nta}\}
\end{array} \right., \quad
Y(u)= \left\{
\begin{array}{lcl}
  1 & \mbox{if} & u=\tb{\ntb} \\
   0 & \mbox{if} & u \in \T\backslash\{\tb{\ntb}\}
\end{array} \right..
\end{eqnarray}
is a free Lie algebra over the set $\{X,Y\}$ \cite{murua06tha}. 
In what follows, we denote as $\L(X,Y)$ the Lie subalgebra of $\g$ generated by the maps $X$ and $Y$. 

 It has also been shown in  \cite{murua06tha} that for each particular Hall--Viennot basis 
$\{E_i \ : \ i=1,2,3,\ldots \}$, (whose elements are given by
(\ref{eq:Ei}) for appropriate values of $i',i''<i$, $i=3,4,\ldots$, and $X$ and $Y$ given 
by (\ref{eq:XY})) \emph{one can associate a bicoloured rooted
 tree $u_i$ to each element $E_i$} such that,
 for any map $\alpha \in \L(X,Y)$, 
\begin{eqnarray}  
  \label{eq:alphaXY}
\alpha &=& \sum_{i\geq 1} \frac{\alpha(u_i)}{\sigma(u_i)} E_i,
\end{eqnarray}
where for each $i$, $\sigma(u_i)$ is certain positive integer associated to the bicoloured rooted tree $u_i$ (the number of symmetries of $u_i$, that we call {\em symmetry number of $u_i$}). For instance, the bicoloured rooted trees $u_i$ and their symmetry numbers $\sigma(u_i)$ associated to the elements $E_i$ (of degree $|i|\leq 5$) of the Hall basis used in this work are displayed in Table~\ref{tab:Eiui}.

As in the introduction, 
we denote by $\L(X,Y)_n$ ($n\geq 1$) the homogeneous subspace of 
$\L(X,Y)$ of degree $n$ (whence admiting $\{E_i \ : \ |i|=n\}$ as a basis).
% (i.e., the   subspace of $\L(X,Y)$ spanned by Lie polynomials of degree $m$).  ), so that $\L(X,Y) = \L(X,Y)_1 \oplus \L(X,Y)_2 \oplus \cdots$.
It can be seen~\cite{murua06tha} that, if  $\alpha \in \L(X,Y)$, then its projection $\alpha_n$ to the homogeneous subspace $\L(X,Y)_n$ is given by
\begin{eqnarray}
\label{eq:alpha_n}
 \alpha_n(u)= \left\{
\begin{array}{lcl}
  \alpha(u) & \mbox{if} & |u|=n \\
   0 & \mbox{otherwise} & 
\end{array} \right.
\end{eqnarray}
for each $u \in \T$. 

We also use the notation $\overline{\L(X,Y)}$ for the Lie algebra of Lie series, that is, series of the form
 \begin{eqnarray*}
   \alpha = \alpha_1 + \alpha_2 + \alpha_3 + \cdots, \quad \mbox{where} \quad \alpha_n \in \L(X,Y)_n.
 \end{eqnarray*}
Notice that in this setting, a Lie series $\alpha \in \overline{\L(X,Y)}$ is a map $\alpha:\T\to \R$ satisfying that, for each $n\geq 1$, the map $\alpha_n$ given by (\ref{eq:alpha_n}) belongs to $\L(X,Y)_n$. 
A map $\alpha \in \g$ is then a Lie series if and only if (\ref{eq:alphaXY}) holds (see 
\cite{murua06tha} for an alternative characterization of maps $\alpha:\T \rightarrow \R$ 
that actually belong to $\overline{\L(X,Y)}$).

 In particular, the BCH series $Z = Z_1 + Z_2 + Z_3 + \cdots$ given by
 (\ref{varada}) (for $X$ and $Y$ defined as in (\ref{eq:XY})) is a Lie series. From (\ref{varada}), it follows that
 $Z(\tb{\nta})=Z(\tb{\ntb})=1$, and for $n=2,3,4,\ldots$ 
 \begin{eqnarray}
\label{varada2}
         n Z(u) & = & \frac{1}{2} [X-Y, Z](u)  + \sum_{p=1}^{[(n-1)/2]} \frac{B_{2p}}{(2p)!} \left(\ad_{Z}^{2p} (X+Y)\right)(u)
 \end{eqnarray}
for each
 $u \in \T$ with $n=|u|$. 
 Recall that, for arbitrary $\alpha,\beta \in \g$ and $u \in \T$, the
 value $[\alpha,\beta](u)$ is defined in terms of bicoloured rooted
 trees $u_{(j)},u^{(j)}$ with less vertices than $u$, so that
 (\ref{varada2}) effectively allows us to compute the values $Z(u)$ for
 all bicoloured rooted trees with arbitrarily high number $|u|$ of
 vertices. 
In this way, the characterization (\ref{eq:alphaXY}) of maps $\alpha \in \g$ that are Lie series directly gives a way to write $Z \in \overline{\L(X,Y)}$ in the form (\ref{eq:BCHEi}) with 
\begin{eqnarray}
\label{eq:BCHtrees}
  z_i = \frac{Z(u_i)}{\sigma(u_i)} \quad \mbox{for} \quad i\geq 1.
\end{eqnarray}
For instance, we have according to Table~\ref{tab:Eiui} that in the Hall basis,
\begin{eqnarray*} 
  Z &=&   \sum_{i\geq 1} z_i  E_i\ = \ 
  \sum_{i\geq 1} \frac{Z(u_i)}{\sigma(u_i)} E_i\\
&=&  Z(\tb{\nta}) X + Z(\tb{\ntb}) Y + Z(\tb{\nte}) [Y,X] 
\\  && 
+  \frac{Z(\tb{\ntah})}{2} [[Y,X],X] +
Z(\tb{\ntai}) [[Y,X],Y] + \cdots,
\end{eqnarray*}
where the first five coefficients $Z(u_i)$ can be obtained 
by applying (\ref{varada2}) with (\ref{eq:[a,b]ex}):
\begin{eqnarray}
  [X-Y,Z](\tb{\nte}) &=& -Z(\tb{\nta}) - Z(\tb{\ntb}) \ = \ -2, \nonumber \\
  2 Z(\tb{\nte}) &=& \frac{1}{2}  [X-Y,Z](\tb{\nte}) \ = \ -1, \nonumber \\
  {[X-Y,Z]}(\tb{\ntf}) &=& 0,  \quad 2 Z(\tb{\ntf}) \ = \ 0,\nonumber \\ 
  {[X-Y,Z]}(\tb{\ntah}) &=& -2 Z(\tb{\nte}) \ = \ 1, \nonumber \\
  {[X-Y,Z]}(\tb{\ntai}) &=& Z(\tb{\nte})-Z(\tb{\ntf}) \ = \ -\frac{1}{2}, \label{eq:Z(u)ex} \\
{[Z,[Z,X+Y]]}(\tb{\ntah}) &=& -2 Z(\tb{\nta}) (Z(\tb{\nta})-Z(\tb{\ntb})) \ = \ 0, \nonumber \\
{[Z,[Z,X+Y]]}(\tb{\ntai}) &=&  -Z(\tb{\nta}) (Z(\tb{\nta})-Z(\tb{\ntb})) \ = \ 0, \nonumber \\
3 Z(\tb{\ntah}) &=& \frac{1}{2}, \nonumber \\
3 Z(\tb{\ntai}) &=& -\frac{1}{4}. \nonumber
\end{eqnarray}
 In summary, the idea of the formalism is to construct algorithmically a sequence of labeled
 rooted trees in a one-to-one correspondence with a Hall basis, verifying in addition 
 (\ref{eq:alphaXY}). In this way it is quite straightforward to build and characterize Lie series, and
 in particular, the BCH series.
 
 \subsection{Detailed treatment}

In this subsection we provide a detailed treatment of the main steps involved in the
procedure previously sketched, first by analyzing the 
representation (\ref{eq:alphaXY}) of Lie series for the classical Hall basis and then by
considering Hall--Viennot bases.

We start by providing an algorithm that constructs the 
table of values $(i',i'')$ (for $i\geq 3$) in (\ref{eq:Ei}) (together with $|i|$ for $i\geq 1$)  
that determines a classical Hall basis. The algorithm starts by setting 
\begin{eqnarray*}
  1' = 1, \quad 1''=0, \quad 2'=2, \quad 2''=0, \quad |1|=1, \quad |2|=1,
\end{eqnarray*}
and initializing the counter $i$ as $i=3$. Then, 
the values $i',i'',|i|$ for subsequent values of $i$ are set as follows
($i^{++}$ indicates that the value of the counter $i$ is incremented by one): 
\begin{algorithm}
\label{alg:PHall}
\begin{eqnarray*}
   \begin{array}{cclclclcl}
\mbox{\textit{for}}  \;\;\; n &\!=\!& 2,3,\ldots                      \\
    &     & j = 1,\ldots,i-1                \\
    &     &      \qquad  k = j+1, \ldots,i-1 \\
    &     &       \qquad \qquad \mbox{If} \quad |j|+|k|=n \mbox{ and } j\geq k'' \quad
\mbox{then} \\
    &     &     \qquad \qquad \phantom{\mbox{If}} \quad i'' = j, \ i'=k, \ |i|=n, \\
    &     &     \qquad \qquad \phantom{\mbox{If}} \quad i^{++}.
  \end{array}
\end{eqnarray*}
\end{algorithm}
The values of $i'$, $i''$, $|i|$ thus determined satisfy that 
$i'>i''\geq (i')''$ for $i\geq 3$.
In addition,  $j<i$ if $|j|<|i|$, which implies that $i',i''<i$ for all $i\geq 3$.
The values for $|i|$, $i'$, and $i''$ and the element $E_i$  of the basis
 for the values of the index $i$ of degree $|i|\leq 5$ are displayed in
Table~\ref{tab:Eiui}.

%The algorithm above has been written for clarity. More efficient versions of it can be construted by setting, together with $i'',i',|i|$, the values $f_n$ such that
%\begin{eqnarray}
%%
%\label{eq:cond02}
%  \{i \in \Z \ : \ |i|=n\} = \{f_n, f_{n}+1, \ldots, f_{n+1}-1\}.
%\end{eqnarray}
%%

On the other hand, it is possible to design a simple recursive procedure to define the
bicoloured rooted trees $u_i$ appearing in (\ref{eq:alphaXY}) in terms of the values of $i'$ and $i''$
by using the following binary operation. 
Given $u,v\in \T$, the new rooted tree $u \cir v \in \T$ is a rooted tree with $|u|+|v|$ vertices
obtained by grafting the rooted tree $v$ to the root of $u$ (that is to say, $u \cir v$ is a new bicoloured rooted tree with the coloured vertices of $u$ and $v$, one edge that makes the root of $v$ a child of the root of $u$ added to the edges of $u$ and $v$).  For instance,
\begin{eqnarray*}
\tb{\ntah} \cir \tb{\nte} =  \tb{\ntbfg}, \qquad \mbox{ and also } \qquad
\tb{\ntei} \cir \tb{\nta} =  \tb{\ntbfg}.
\end{eqnarray*}
We now define 
 \begin{eqnarray}
   \label{eq:u_i}
u_1=\tb{\nta}, \quad u_2=\tb{\ntb}, \quad \mbox{and} \quad u_i = u_{i'} \cir u_{i''} \quad \mbox{for} \quad i\geq 3.
 \end{eqnarray}
Finally, the symmetry numbers $\sigma_i=\sigma(u_i)$ can also be determined recursively:
\begin{eqnarray}
  \label{eq:sigma_i}
\sigma_1=\sigma_2=1, \qquad \mbox{ and } \qquad \sigma_i = \kappa_i \, \sigma_{i'} \sigma_{i''}, \quad \mbox{for} \quad i\geq 3,
\end{eqnarray}
where $\kappa_i=1$ if $(i')''\neq i''$, and 
$\kappa_i=\kappa_{i'}+1$ if $(i')''=i''$.

The bicoloured rooted trees $u_i$, their symmetry numbers $\sigma(u_i)$, and the coefficients $z_i = Z(u_i)/\sigma(u_i)$ in the BCH series (\ref{eq:BCHEi}) are displayed in Table~\ref{tab:Eiui} for the first values of the index $i$, whereas in Table~\ref{tab:BCH} given in the Appendix, the terms of the BCH series (\ref{eq:BCHEi}) up to terms of degree 9 are given in compact form for the classical Hall basis by displaying the values of $i'$, $i''$, and $z_i = Z(u_i)/\sigma(u_i)$ for each index $i$.

\begin{table}[t]
\renewcommand{\arraystretch}{1.2}
  \centering
  \begin{tabular}{|c|c|c|c|c|c|c|c|c|} \hline
$i$ & $|i|$ & $i'$ & $i''$ & $E_i$ & $u_i$ & $\sigma(u_i)$ & $z_i = \frac{Z(u_i)}{\sigma(u_i)}$\\ \hline
1 & 1  &  & & $X$ & $\tb{\nta}$ & 1 & 1\\
2 & 1  & & & $Y$ & $\tb{\ntb}$ & 1 & 1\\ \hline
3 & 2  & 2 & 1 & $[Y,X]$ & $\tb{\nte}$ & 1 & $ -\frac{1}{2}$ \\ \hline
4 & 3  & 3 & 1 & $[[Y,X],X]$ & $\tb{\ntah}$ & 2 & $ \frac{1}{12}$ \\
5 & 3  & 3 & 2 & $[[Y,X],Y]$ & $\tb{\ntai}$ & 1 &  $-\frac{1}{12}$\\  \hline
6 & 4  & 4 & 1 & $[[[Y,X],X],X]$ & $\tb{\ntfi}$ & 6 & 0 \\
7 & 4  & 4 & 2 & $[[[Y,X],X],Y]$ & $\tb{\ntgo}$ & 2 & $ \frac{1}{24}$\\
8 & 4  & 5 & 2 & $[[[Y,X],Y],Y]$ & $\tb{\ntga}$ & 2 & 0 \\ \hline
9 & 5  & 6 & 1 & $[[[[Y,X],X],X],X]$ & $\tb{\ntbhb}$ & 24 &  $-\frac{1}{720}$ \\
10 & 5 & 6 & 2 & $[[[[Y,X],X],X],Y]$ & $\tb{\ntbhc}$ & 6 &  $-\frac{1}{180}$\\
11 & 5 & 7 & 2 & $[[[[Y,X],X],Y],Y]$ & $\tb{\ntbhd}$ & 4 & $ \frac{1}{180}$\\
12 & 5 & 8 & 2 & $[[[[Y,X],Y],Y],Y]$ & $\tb{\ntbhe}$ & 6 & $ \frac{1}{720}$\\
13 & 5 & 4 & 3 & $[[[Y,X],X],[X,Y]]$ & $\tb{\ntbfg}$ & 2 &  $-\frac{1}{120}$\\
14 & 5 & 5 & 3 & $[[[Y,X],Y],[X,Y]]$ & $\tb{\ntbga}$ & 1 &  $-\frac{1}{360}$\\
\hline
  \end{tabular}
  \caption{First elements $E_i$ of the basis of P. Hall, their corresponding bicoloured rooted trees $u_i$, the values $|i|$, $i''$, $i'$, $\sigma(u_i)$, and the coefficients $z_i=Z(u_i)/\sigma(u_i)$ in the BCH series (\ref{eq:BCHEi})}
  \label{tab:Eiui}
\end{table}

This procedure can be extended indeed to Hall--Viennot bases. 
A set $\{E_i  :  i=1,2,3,\ldots\} \subset \L(X,Y)$ recursively defined as (\ref{eq:Ei}) 
with some positive integers $i',i''<i$ ($i=3,4,\ldots$) is a Hall--Viennot basis if there exists a total order relation $\succ$ in the set of indices $\{1,2,3,\ldots\}$
such that $i\succ i''$ for all $i\geq 3$, and the map
  \begin{eqnarray}
    \label{eq:dmap}
    d:\{3,4,\ldots\} &   \longrightarrow &
\{(j,k) \in \Z \times \Z \ : \ j\succ k \succeq j''\}, \quad  \\
d(i)=(i',i'')  &  & 
  \end{eqnarray}
(with the convention $1''=2''=0$) is bijective.

In~\cite{viennot78adl,reutenauer93fla}, Hall--Viennot bases are indexed by a subset of words (a Hall set of words) on the alphabet $\{x,y\}$. Such Hall set of words $\{w_i\ : \ i\geq 1\}$ can be obtained by defining recursively $w_i$ as the concatenation $w_{i'}w_{i''}$ of the words $w_{i'}$ and $w_{i''}$, with $w_1=x$ and $w_2=y$.
For instance, the Hall set of words $w_i$ associated to the indices $i=1,2,\dots,14$ in Table~\ref{tab:Eiui} are $x$, $y$, $yx$, $yxx$, $yxy$, $yxxx$, $yxxy$, $yxyy$, $yxxxx$, $yxxxy$, $yxxyy$, $yxyyy$, $yxxyx$, $yxyyx$.

For the classical Hall basis we have considered before, the map (\ref{eq:dmap}) has been constructed in such a way that the total order relation $\succ$ is the natural order relation in $\Z$, i.e., $>$ (notice that in 
\cite{bourbaki89lga} the total order is chosen as $<$).

This is not possible, however, for the Lyndon basis. 
The Lyndon basis can be constructed as a Hall--Viennot basis by considering the order relation $\succ$ as follows: $i \succ j$ if, in lexicographical order (i.e., the order used when ordering words in the dictionary), the Hall word $w_i$ associated to $i$ comes before than the Hall word $w_j$ associated to $j$. The Hall set of words $\{w_i\ : \ i\geq 1\}$ corresponding to the Lyndon basis is the set of Lyndon words, which  can be defined as the set of words $w$ on the alphabet $\{x,y\}$ satisfying that, for arbitrary decompositions of $w$ as the concatenation $w=u v$ of two non-empty words $u$ and $v$, the word $w$ is smaller than $v$ in lexicographical order
\cite{viennot78adl,lothaire83cow}.

Now, the representation (\ref{eq:alphaXY}) of a map $\alpha \in \overline{\L(X,Y)}$ (and in particular, the BCH series (\ref{eq:BCHEi}) with (\ref{eq:BCHtrees})) for any Hall--Viennot
 basis can be stated as follows.

\begin{theorem}
  \label{th:LieSeries}
Given a total order relation $\succ$ in $\Z$ and a bijection (\ref{eq:dmap}) 
satisfying that $i\succ i''$ for all $i\geq 3$, then any map $\alpha \in \overline{\L(X,Y)}$ admits the representation (\ref{eq:alphaXY}) for the Hall basis (\ref{eq:Ei}) and  the bicoloured rooted trees $u_i$ and their symmetry numbers  
$\sigma_i=\sigma(u_i)$ recursively defined as
(\ref{eq:u_i}) and (\ref{eq:sigma_i}).
\end{theorem}
Theorem~\ref{th:LieSeries} can be proven as a corollary of Theorem~3 and Remark 17 in \cite{murua06tha}. Actually, in \cite{murua06tha} it is shown that (\ref{eq:alphaXY}) holds for a different set $\hat \T = \{u_1,u_2,u_3,\ldots\}$ of bicoloured rooted trees associated to a Hall basis, for which $\sigma(u_i)=1$ for all $i$.  However, the set of Hall rooted trees we consider here (which is 
the set $\hat \T^*$ considered in Remark 17 in \cite{murua06tha}) has some advantages from the 
computational point of view.

In Table~\ref{tab:EiuiLyndon}, we display the elements $E_i$ of the Lyndon basis with degree $|i|\leq 5$, the corresponding Lyndon words $w_i$, the bicoloured rooted trees $u_i$, the values $|i|$, $i''$, $i'$, $\sigma(u_i)$, and the coefficients $z_i=Z(u_i)/\sigma(u_i)$ in the BCH series (\ref{eq:BCHEi}).

\begin{table}[t]
\renewcommand{\arraystretch}{1.2}
  \centering
  \begin{tabular}{|c|c|c|c|c|c|c|c|c|} \hline
$i$ & $|i|$ & $i'$ & $i''$ & $w_i$ & $E_i$ & $u_i$ & $\sigma(u_i)$ & $z_i=\frac{Z(u_i)}{\sigma(u_i)}$\\ \hline
1 & 1 &  & & $x$ & $X$ & $\tb{\nta}$ & 1 & 1\\
2 & 1 & & & $y$ & $Y$ & $\tb{\ntb}$ & 1 & 1\\ \hline
3 & 2 & 1 & 2 & $xy$ & $[X,Y]$ & $\tb{\ntd}$ & 1 & $ \frac{1}{2}$ \\ \hline
4 & 3 & 3 & 2 & $xyy$ & $[[X,Y],Y]$ & $\tb{\ntag}$ & 2 & $ \frac{1}{12}$ \\
5 & 3 & 1 & 3 & $xxy$ & $[X,[X,Y]]$ & $\tb{\nth}$ & 1 &  $\frac{1}{12}$\\  \hline
6 & 4 & 4 & 2 & $xyyy$ & $[[[X,Y],Y],Y]$ & $\tb{\ntfh}$ & 6 & 0 \\
7 & 4 & 1 & 4 & $xxyy$ & $[X,[[X,Y],Y]]$ & $\tb{\ntca}$ & 2 & $ \frac{1}{24}$\\
8 & 4 & 1 & 5 & $xxxy$ & $[X,[X,[X,Y]]]$ & $\tb{\ntbb}$ & 1 & 0 \\  \hline
9 & 5 & 6 & 2 & $xyyyy$ & $[[[[X,Y],Y],Y],Y]$ & $\tb{\ntbha}$ & 24 &  $\frac{1}{720}$ \\
10 & 5 & 5 & 3 & $xxyxy$ & $[[X,[X,Y]],[X,Y]]$ & $\tb{\ntbcg}$ & 2 &  $\frac{1}{360}$\\
11 & 5 & 3 & 4 & $xyxyy$ & $[[X,Y],[[X,Y],Y]]$ & $\tb{\ntboa}$ & 2 & $ \frac{1}{120}$\\
12 & 5 & 1 & 6 & $xxyyy$ & $[X,[[[X,Y],Y],Y]]$ & $\tb{\ntabo}$ & 6 & $ \frac{1}{180}$\\
13 & 5 & 1 & 7 & $xxxyy$ & $[X,[X,[[X,Y],Y]]]$ & $\tb{\nthc}$ & 2 &  $\frac{1}{180}$\\
14 & 5 & 1 & 8 & $xxxxy$ & $[X,[X,[X,[X,Y]]]]$ & $\tb{\ntgd}$ & 1 &  $-\frac{1}{720}$\\
\hline
  \end{tabular}
  \caption{First elements $E_i$ of the Lyndon basis, their corresponding Lyndon words $w_i$ and bicoloured rooted trees $u_i$, the values $|i|$, $i''$, $i'$, $\sigma(u_i)$, and the coefficients $z_i=Z(u_i)/\sigma(u_i)$ in the BCH series (\ref{eq:BCHEi})}.
 \label{tab:EiuiLyndon}
\end{table}

\subsection{Practical aspects in the implementation}

An important ingredient in the whole procedure is the practical implementation 
of the Lie bracket $[\alpha,\beta]$ of two Lie series $\alpha,\beta \in \overline{\L(X,Y)} \subset \g$,
which we address next. Let us consider for each $u \in \T$ the sequence 
\begin{eqnarray}
\label{eq:S(u)def}
  S(u) = \{(u_{(1)},u^{(1)}),\ldots,(u_{(|u|-1)},u^{(|u|-1}))\}
\end{eqnarray}
  of pairs of bicoloured rooted trees used to define the Lie bracket $[\alpha,\beta]$ in (\ref{eq:bracket}). For instance, 
\begin{eqnarray*}
 S( \tb{\ntdf} ) = \left\{(\tb{\ntah},\tb{\ntb}), (\tb{\nta},\tb{\ntac}), (\tb{\nta},\tb{\ntac}) \right\}.
\end{eqnarray*}
It can be seen that the sequences $S(u)$ satisfy the following recursion. If $u = v \circ w$, where $v,w \in \T$, then, let $p=|v|-1$, $q=|w|-1$, and 
\begin{eqnarray*}
  S(v) =  \{(v_{(1)},v^{(1)}), \ldots, (v_{(p)},v^{(p)})\}, \quad
  S(w) =  \{(w_{(1)},w^{(1)}), \ldots, (w_{(q)},w^{(q)})\},
\end{eqnarray*}
then
\begin{equation}
  \label{eq:S(u)}
S(u) = \{(w,v),(v_{(1)},v^{(1)}\circ w), \ldots, (v_{(p)}, v^{(p)} \circ w),
(w_{(1)}, v \circ w^{(1)}), \ldots, (w_{(q)},v \circ w^{(q)})\}.
\end{equation}
From the point of view of implementation, it is important to observe that, if one wants to compute the BCH formula (\ref{eq:BCHEi}) up to terms of degree $|i|\leq n$, there is no need to compute all values $Z(u)$ for $u \in T$ with $|u|\leq n$, since only the values (\ref{eq:BCHtrees}) of $Z(u)$ for the rooted trees $u_i$ associated to the elements $E_i$ of the basis are used in (\ref{eq:BCHEi}). However, when applying the recursion (\ref{varada2}) for the bicoloured rooted trees $u_i$, $i=1,2,3,\ldots$ (that we will call Hall rooted trees), one finds that $Z(u)$ needs to be computed for some additional bicoloured rooted trees $u$.  Observe for instance that in (\ref{eq:Z(u)ex}), the value $Z(u)$ for $u=\tb[-0.1]{\ntf}$ was needed in order to get the value $Z(\tb{\ntai})$ from the recursion (\ref{varada2}). This is due to the fact that in (\ref{eq:bracket}), the bicoloured rooted trees $u_{(j)}$ need not be Hall rooted trees when $u$ is a Hall rooted tree ($u^{(j)}$ is however necessarily a Hall rooted tree in that case). 

The minimal set $\tilde \T_n$ of bicoloured rooted trees $u$ for which
$Z(u)$ needs to be computed in order to get the values of $Z(u_i)$ for
Hall rooted trees with $|i|\leq n$ by using recursion (\ref{varada2})
can be determined by requiring that 
\begin{eqnarray*}
  \{u_i \ :\ |i|\leq n\} \subset \tilde \T_n \subset \T
\quad \mbox{and} \quad
 S(\tilde \T_n) \subset \tilde \T_n \times
\tilde \T_n.
\end{eqnarray*}
 It can be seen that the subset $\tilde \T_n$ of bicoloured rooted trees can be alternatively defined as follows: We say that a bicoloured rooted tree $v \in \T$  is covered by $u \in \T$ if either $v$ can be obtained from $u$ by removing some of its vertices and edges, or  $u=v$. For instance, the bicoloured rooted trees covered by the tree $u_{11}$ in Table~\ref{tab:EiuiLyndon} are
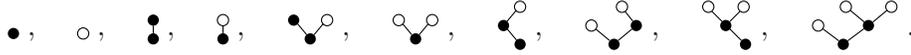
\begin{eqnarray*}
  \tb{\nta}, \quad \tb{\ntb}, \quad \tb{\ntc}, \quad \tb{\ntd}, \quad \tb{\ntaf}, \quad \tb{\ntag}, \quad \tb{\nth}, \quad \tb{\nted}, \quad \tb{\ntca}, \quad \tb{\ntboa}.
\end{eqnarray*}
Then, it can be seen that $\tilde \T_n$ is the set of bicoloured rooted trees covered by some of the trees in $\{u_i \ :\ |i|\leq n\}$. 

As a summary of this treatment, 
we next describe the main steps of the algorithm that we use to compute the BCH series up to terms of a given degree $N$ for an arbitrary Hall--Viennot basis. Let $m_N$ be sum of the dimensions of the homogeneous subspaces $\L(X,Y)_n$ for $1\leq n \leq N$, and let $\tilde m_N$ be the number of bicoloured rooted trees in $\tilde \T_N$ (so that $m_N\leq \tilde m_N$). We proceed as follows for a given $N$:

\begin{enumerate}
 \item Determine the values $i',i''$ for each $i=1,\ldots,m_N$ such that the $E_i$ given by (\ref{eq:Ei}) are the elements of degree $|i|\leq N$ of the required Hall--Viennot basis. Algorithm~\ref{alg:PHall} can be used in the case of the basis of P. Hall. We use a similar (although slightly more complex) algorithm for the general case.
\item  Determine the bicoloured rooted trees $u \in \tilde T_N$ together with the $|u|-1$ pairs of bicoloured rooted  trees in $S(u)$ recursively obtained by (\ref{eq:S(u)}). Actually, we associate each bicoloured rooted tree in $\tilde \T_N$ with a positive integer, such that
$\tilde \T_N = \{ u_i \ : \ i=1,2,\ldots,\tilde m_N\}$ (and
$ \{ u_i \ : \ i=1,2,\ldots,m_N\}$ is the set of Hall trees of degree $|i|\leq N$). Each $S(u_i)$ is then represented as a list of $|i|-1$ pairs of positive integers.
\item Represent the truncated versions of Lie series $\alpha$ (truncated up to terms of degree $N$) as a list of $\tilde m_N$ real values $(\alpha_1,\ldots,\alpha_{\tilde m_N})$ corresponding to $(\alpha(u_1),\ldots,\alpha(u_{\tilde m_N}))$. The Lie bracket $\gamma=[\alpha,\beta]$ of two Lie series can be implemented as a way to obtain the list $(\gamma_1,\ldots,\gamma_{\tilde m_N})$ from the lists $(\alpha_1,\ldots,\alpha_{\tilde m_N})$ and $(\beta_1,\ldots,\beta_{\tilde m_N})$ in terms of the pairs of integers representing $S(u_i)$ for each $i=1,\ldots,\tilde m_N$.
\item Represent the truncated versions of BCH series $Z$ (truncated up to terms of degree $N$) as a list of $\tilde m_N$ rational values $(Z_1,\ldots,Z_{\tilde m_N})$ corresponding to $(Z(u_1),\ldots,Z(u_{\tilde m_N}))$, which can be obtained by initializing that list as $(1,1,0,\ldots,0)$ and 
applying (\ref{varada2}) repeatedly for $n=2,\ldots,N$.
\end{enumerate}

It is worth noticing that the number of trees in $\tilde \T_n$ is different for different Hall--Viennot bases. For instance, for the basis of P. Hall, $\tilde T_{20}$ has 724018 bicoloured rooted trees, while for the Lyndon basis the set $\tilde T_{20}$ has 1952325 bicoloured rooted trees. Due to this fact, 
the amount of memory and CPU time required to compute with our algorithm the BCH formula up to a given degree for the Lyndon basis is considerably larger than for the basis of P. Hall. Moreover, the number of non-zero coefficients $z_i$ in the BCH formula differs considerably in both bases. For instance, there are 109697 non-vanishing coefficients $z_i$ (out of $111013$ elements $E_i$ of degree $|i|\leq 20$) in the BCH formula for the basis of P. Hall, while for the Lyndon basis the number of non-vanishing coefficients $z_i$ is $76760$.

\section{Optimal convergence domain of the BCH series}

\subsection{The BCH formula and the Magnus expansion}
  \label{conv.3.2}

One particularly simple way of obtaining a sharp bound on the convergence domain 
for the BCH series consists in relating it
with the Magnus expansion for linear differential equations.  For the sake of completeness, we summarize here the main features of this procedure.  

Suppose we have the non-autonomous linear differential equation
\begin{equation}   \label{mag.1.1}
    \frac{dU}{dt} = A(t) U, \qquad U(0)=I,
\end{equation}
where $U(t)$ and $A(t)$ are operators acting on some Hilbert space $\mathcal{H}$
(in particular,  $n \times n$ real or complex matrices). Then the idea is to express the 
solution $U(t)$ as the exponential of a certain operator $\Omega(t)$, 
\begin{equation}   \label{mag.1.2}
   U(t) = \exp \Omega(t).
\end{equation}
By substituting (\ref{mag.1.2}) into (\ref{mag.1.1}), one can derive the differential equation
satisfied by the exponent $\Omega$:
\begin{equation}   \label{mag.1.3} 
  \Omega^{\prime} =  \sum_{k=0}^{\infty} \frac{B_k}{k!}
  \mathrm{ad}_{\Omega}^k (A(t)) , \qquad \Omega(0) = O.
\end{equation}
By applying Picard's iteration on
(\ref{mag.1.3}), one gets an infinite series for $\Omega(t)$,
\begin{equation}   \label{mag.1.4}
   \Omega(t) = \sum_{m=1}^{\infty} \Omega_m(t),
\end{equation}     
whose terms can be obtained recursively from
\begin{eqnarray}   \label{mag.1.5b}
    \Omega_{1}(t)  & = & \int_{0}^{t}A(t_{1})~dt_{1},\nonumber\\   
    \Omega_m(t)   &  = &  \sum_{j=1}^{m-1} \frac{B_j}{j!} \,
        \int_0^t \,
       (\ad_{\Omega(s)}  A(s))_m \, ds    \qquad m \ge 2.
\end{eqnarray}
Equations (\ref{mag.1.2}) and (\ref{mag.1.4}) constitute the so-called \emph{Magnus
expansion} for the solution of (\ref{mag.1.1}), whereas the infinite series (\ref{mag.1.4}) 
with (\ref{mag.1.5b}) is
known as the \emph{Magnus series}.

Since the 1960s \cite{weiss62tbf}, 
the Magnus expansion has been successfully applied as a perturbative
tool in numerous areas of physics and chemistry, from atomic and molecular
physics to nuclear magnetic resonance and quantum electrodynamics (see
\cite{blanes98maf}  and \cite{blanes07mem} for a review and a list of references). 
Also, since the work
by Iserles and N{\o}rsett \cite{iserles99ots}, it has been used as a tool to construct
practical algorithms for the numerical integration of equation (\ref{mag.1.1}), while
preserving the main qualitative properties of the exact solution. 

In general, the Magnus series does not converge unless $A$ is small in a suitable sense,
and several bounds to the actual radius of convergence have been obtained along the
years. Recently, the following theorem has been proved  \cite{casas07scf}:
\begin{theorem}  \label{conv-mag}
  Let us consider the differential equation $U' = A(t) U$ defined in 
  a Hilbert space $\mathcal{H}$, $\mathrm{dim} \, \mathcal{H} < \infty$,
   with $U(0)=I$, and let $A(t)$ be a bounded linear operator
  on $\mathcal{H}$. Then, the Magnus series 
  $\Omega(t) = \sum_{k=1}^{\infty}  \Omega_k(t)$, with
$\Omega_k$ given by (\ref{mag.1.5b}) converges  in the interval $ t \in [0,T)$  such that
\[
   \int_0^T \|A(s)\| \, ds < \pi
\]
and the sum $\Omega(t)$ satisfies $\exp \Omega(t) = U(t)$. The statement also holds when
$\mathcal{H}$ is infinite-dimensional if $U$ is a normal operator (in particular, if $U$ is
unitary). Here $\| \cdot \|$ stands for the norm defined by the inner product on $\mathcal{H}$.
\end{theorem}    

Moreover, it has been shown that the convergence domain of the Magnus series 
provided by this theorem
is the best result one can get for a generic bounded operator $A(t)$ in a Hilbert
space, in the sense that it is possible to find specific $A(t)$ where
the series diverges for any time $t$ such that  $\int_0^t \|A(s)\| ds > \pi$ 
\cite{moan02obe,casas07scf}.

Now, given two operators $X$ and $Y$, let us consider equation (\ref{mag.1.1}) with
  \begin{equation}    \label{mag.1.6}
   A(t) = \left\{  \begin{array}{ccl}
                   Y &   &   0 \le t <  1  \\
                   X &   &   1 \le t \le 2
                    \end{array}  \right.
\end{equation}
Clearly, the exact solution at $t=2$ is given by $U(2) = \e^{X} \, \e^Y$. On the other
hand, if we apply recurrence (\ref{mag.1.5b}) to compute $U(2)$ with the  Magnus 
expansion, $U(2) = \e^{\Omega(2)}$, we get $\Omega_1(2) = X + Y$ and more generally
$\Omega_n(2) = Z_n$ in (\ref{eq.1.3}). In other words, the BCH series can be considered 
as the Magnus expansion corresponding to the differential equation (\ref{mag.1.1})
with $A(t)$ given by (\ref{mag.1.6}) at $t=2$. 

Since $\int_0^{t=2} \|A(s)\| ds = \|X\| + \|Y\|$, Theorem \ref{conv-mag} leads to the following  bound on
the convergence of the BCH series.
\begin{theorem}   \label{conv-cbh}
  Let $X$ and $Y$ be two bounded elements in a Hilbert space $\mathcal{H}$ with
  $\mathrm{dim } \,  \mathcal{H} \ge 2$.Then the  BCH formula in the form 
  (\ref{eq.1.3}), i.e., expressed as a series of homogeneous
  Lie polynomials in $X$ and $Y$, converges when $\|X\| + \|Y\| < \pi$.
\end{theorem}
Of course, this result can be generalized to any set $X_1, X_2, \ldots, X_k$ of bounded operators:
the corresponding BCH series is convergent if $\|X_1\| + \cdots + \|X_k\| < \pi$ in the $2$-norm.

Let us illustrate the result provided by Theorem \ref{conv-cbh} with a 
simple example involving $2 \times 2$ matrices. \\

\noindent \textbf{Example 1}. Given 
\begin{equation}   \label{exam.1}
    X =   \left(
  \begin{array}{cr}
  \alpha  & 0 \\
  0  &  -\alpha
  \end{array} \right), \qquad
  Y =   \left(
  \begin{array}{cc}
  0  & \beta \\
  0  & 0
  \end{array} \right)
\end{equation}
with $\alpha, \beta \in \mathbb{C}$, a simple calculation shows that
\[
   \log(\e^{ X} \e^{ Y})  = X +
    \frac{2\alpha}{1- \e^{-2\alpha}} \, Y,
\]
which is an analytic function for $|\alpha| < \pi$ with first singularities
at $\alpha = \pm i \pi$. Therefore, the BCH formula cannot converge if 
$|\alpha| \ge \pi$, independently of $\beta \ne 0$. By taking the spectral norm, it is clear that
$\|X\| = |\alpha|$, $\|Y\| = |\beta|$, so that the convergence domain given by Theorem \ref{conv-cbh} 
is $|\alpha| + |\beta| < \pi$.  Notice that in the limit
$|\beta| \rightarrow 0$ this domain is optimal. $\Box$ 

Generally speaking, however, the bound given by Theorem \ref{conv-cbh}
is conservative, i.e., the BCH series converges 
for larger values of $\|X\|$ and $\|Y\|$. Thus, in the previous example, for any $\alpha$ and $\beta$
with $|\alpha| < \pi$ and $|\alpha| + |\beta| \ge \pi$, the BCH series also converges. One would
like therefore to have a more realistic characterization  of this feature. It turns out that this is
indeed feasible for complex $n \times n$ matrices.

\subsection{Convergence for matrices}
   \label{conv.3.3}

\subsubsection{Convergence determined by the eigenvalues}
 \label{cdeeig}

For complex $n \times n$ matrices it is possible to use the theory of analytic matrix functions 
and more specifically, the
logarithm of an analytic matrix function, in a similar way as in the Magnus expansion
\cite{casas07scf}, to characterize more precisely the convergence of the BCH series.

To begin with, let us introduce a parameter $\varepsilon \in \mathbb{C}$ and consider the
substitution $(X,Y) \longmapsto (\varepsilon X, \varepsilon Y)$ into eq. (\ref{neq.1}). It is clear that
\[
  U(\varepsilon) \equiv  \e^{\varepsilon X} \, \e^{\varepsilon Y} 
\]
is an analytic function of $\varepsilon$, $\det U(\varepsilon) \ne 0$ and the matrix function
$Z(\varepsilon) = \log U(\varepsilon)$ is also analytic at $\varepsilon = 0$. Equivalently,
the series $Z(\varepsilon)$ is convergent for sufficiently small $\varepsilon$. 
It turns out that the actual radius 
of convergence of this series is related with the existence of multiple eigenvalues of 
$U(\varepsilon)$. Let us denote by $\rho_1(\varepsilon), \ldots, \rho_n(\varepsilon)$ the
eigenvalues of the matrix $U(\varepsilon)$. Observe that $U(0) = I$, so that 
$\rho_1(0) = \cdots = \rho_n(0) = 1$ and we can take the principal values of the logarithm,
$ \log \rho_1(0) = \cdots = \log \rho_n(0) = 0$.
In essence, if the analytic matrix function $U(\varepsilon)$
has an eigenvalue $\rho_0(\varepsilon_0)$ of multiplicity $l > 1$ 
for a certain $\varepsilon_0$ such that: (a) there is a curve in the $\varepsilon$-plane
joining $\varepsilon = 0$ with $\varepsilon = \varepsilon_0$, and
(b) the number of equal terms in 
$\log \rho_1(\varepsilon_0)$, $\log \rho_2(\varepsilon_0)$, $\ldots, 
\log \rho_{l}(\varepsilon_0)$ such that  $\rho_k(\varepsilon_0) = \rho_0$, $k=1,\ldots, l$ is less
than the maximum dimension of the elementary Jordan block corresponding to $\rho_0$, then
the radius of convergence of the series $Z(\varepsilon) = 
\sum_{k \ge 1} \varepsilon^k Z_{k}$ verifying
$\exp Z(\varepsilon) = U(\varepsilon)$ is precisely $r = |\varepsilon_0|$ \cite{casas07scf}. 

More specifically, we find first the values of the parameter $\varepsilon$ for which the 
characteristic polynomial $\det (U(\varepsilon) - \rho I)$ has multiple roots and write them
in order of non-decreasing absolute value,
\begin{equation}  \label{conv.2.1}
   \varepsilon_0^{(1)}, \, \varepsilon_0^{(2)}, \varepsilon_0^{(3)}, \ldots
\end{equation}
Next,  we consider the circle $|\varepsilon| = |\varepsilon_0^{(1)}|$ in the complex 
$\varepsilon$-plane and denote by $\rho_0^{(1)}$ an eigenvalue of 
$U(\varepsilon_0^{(1)})$ with multiplicity $l_1 > 1$. Let $\varepsilon$ move along some fixed curve 
$L$ from $\varepsilon = 0$ to 
$\varepsilon = \varepsilon_0^{(1)}$ in the circle $|\varepsilon| \le |\varepsilon_0^{(1)}|$. Then
it is clear that  $l_1$ eigenvalues $\rho_j(\varepsilon)$ will tend to $\rho_0^{(1)}$ at 
$\varepsilon = \varepsilon_0^{(1)}$. 
If these points lie at 
$\varepsilon = \varepsilon_0^{(1)}$  on the same sheet of the Riemann
surface of the function $\log z$, and this is true for all (possible) multiple eigenvalues 
of $Y_t(\varepsilon)$ at $\varepsilon = \varepsilon_0^{(1)}$, then
$\varepsilon_0^{(1)}$ is called a \emph{extraneous root}. Otherwise, 
$\varepsilon_0^{(1)}$ is called a \emph{non-extraneous root}.

By the analysis carried out in \cite{yakubovich75lde},
when $|\varepsilon| < |\varepsilon_0^{(1)}|$ the numbers $\log \rho_j(\varepsilon)$ are uniquely determined as eigenvalues of the matrix $Z(\varepsilon)$ and this series is convergent. This is
also true at $|\varepsilon| = |\varepsilon_0^{(1)}|$ if $\varepsilon_0^{(1)}$ is an extraneous root,
since then the eigenvalues of $Z(\varepsilon)$ retain their identity throughout the collision process,
so that we proceed to the next value in the
sequence (\ref{conv.2.1}) until
 a non-extraneous root is obtained. 
 
 Assume, for simplicity, that 
$\varepsilon_0^{(2)}$ is the first non-extraneous root, for which there exists an eigenvalue
$\rho_0$ of $U(\varepsilon)$ with multiplicity $l>1$. Associated with
 this multiple eigenvalue $\rho_0$ there is a pair of integers $(p,q)$ defined as
 follows.

The integer $p$ is the greatest number of equal terms in the set of numbers 
$\log \rho_1(\varepsilon_0)$, $\log \rho_2(\varepsilon_0)$, $\ldots, 
\log \rho_{l}(\varepsilon_0)$ such that  $\rho_k(\varepsilon_0) = \rho_0$, $k=1,\ldots, l$. 

The integer $q$ is the maximum degree of the elementary divisors $(\rho - \rho_0)^k$ of 
$U(\varepsilon_0)$,
i.e., the maximum dimension of the elementary Jordan block corresponding to $\rho_0$.

Under these conditions, it has been proved that, if $p < q$ for the eigenvalue $\rho_0$, then
the radius of convergence of the series $Z(\varepsilon) = \sum_{k \ge 1} \varepsilon^k Z_k$
is precisely $r = |\varepsilon_0|$ \cite{yakubovich75lde}.

Although in some cases with $p \ge q$ the series $Z(\varepsilon)$ may converge at 
$|\varepsilon| = |\varepsilon_0|$ and the radius of convergence $r$ is greater than $|\varepsilon_0|$
(for instance, when $X$ and $Y$ are diagonal), this situation is exceptional in a topological sense,
as explained in \cite[pp. 65-66]{yakubovich75lde}.

\subsubsection{Examples}

In order to illustrate this result we next consider a pair of examples involving also $2 \times 2$ matrices. 

\vspace*{0.2cm}

\noindent \textbf{Example 1}. The first example involves again the matrices $X$ and $Y$ given 
by (\ref{exam.1}). In this case
\[
    U(\varepsilon) = \e^{\varepsilon X} \, \e^{\varepsilon Y} =
    \left(  \begin{array}{lc}
        \e^{\varepsilon \alpha}  &  \   \varepsilon \beta \, \e^{\varepsilon \alpha}  \\
                0                                &    \e^{- \varepsilon \alpha}
             \end{array}  \right)
\]                
The first values of $\varepsilon$ for which there are multiple eigenvalues of $U(\varepsilon)$
are
\[
      \varepsilon = 0, \qquad  \varepsilon = \pm i \frac{\pi}{\alpha}.
\]
The first value, $\varepsilon = 0$, is clearly an extraneous root, whereas the eigenvalues of the
matrix $U(\varepsilon)$ move along the unit circle, one clockwise and the other counterclockwise
from 
   \[
       \rho_{1,2}(0) = 1  \quad  \mbox{ to } \quad \rho_{1,2}(i \pi/\alpha) = -1
   \]
when $\varepsilon$ varies along the imaginary axis from $\varepsilon = 0$ to
$\varepsilon = i \pi/\alpha$ (the same considerations apply to the case $\varepsilon = -i \pi/\alpha$). 
Then, obviously, $p=1$ and $q=2$, so that the radius of convergence of the series
$Z(\varepsilon)$ is 
\[
    |\varepsilon| = \frac{\pi}{|\alpha|}. 
\]
By fixing $\varepsilon = 1$, we get the actual domain of convergence of the BCH series as
$|\alpha| = \pi$, i.e., the same result as in section \ref{conv.3.2}. $\Box$ 

\vspace*{0.2cm}

\noindent \textbf{Example 2}. Consider now the matrices
\[
    A =   \left(
  \begin{array}{cr}
  0  & 0 \\
  1  &  0
  \end{array} \right), \qquad
  B =   \left(
  \begin{array}{cc}
  0  & 1 \\
  0  & 0
  \end{array} \right)
\]
and $X= \alpha A$, $Y= \alpha B$, with $\alpha > 0$. Then
\begin{equation}   \label{ex2.1}
    U(\varepsilon) = 
    \left(  \begin{array}{cc}
        1  &  \      \alpha \varepsilon   \\
           \alpha \varepsilon         &   \   1 + \alpha^2 \varepsilon^2
             \end{array}  \right)
\end{equation}                
has multiple eigenvalues when
$\varepsilon_0^{(1)} = 0$, $\varepsilon_0^{(2)} = \pm i \frac{2}{\alpha}$.
As $\varepsilon$ varies along the imaginary axis from $\varepsilon = 0$ to 
$\varepsilon = \varepsilon_0^{(2)}$, the eigenvalues of the matrix $U(\varepsilon)$,
\[
    \rho_{1,2}(\varepsilon) = 1 + \frac{\alpha^2}{2} \varepsilon^2 \pm 
     \sqrt{\left(1 + \frac{\alpha^2}{2} \varepsilon^2 \right)^2 - 1}
\]     
move along the unit circle, one clockwise and the other counterclockwise from
   \[
       \rho_{1,2}(0) = 1  \quad \mbox{ to } \quad \rho_{1,2}(\varepsilon_0^{(2)}) = -1
   \]
Thus, $\rho_1(\varepsilon_0^{(2)})$ and $\rho_2(\varepsilon_0^{(2)})$ lie on different sheets of
  the Riemann surface of the function $\log z$ and therefore $\varepsilon_0^{(2)}$ is a non-extraneous
  root, with $p=1$. Since $U(\varepsilon_0^{(2)})  \ne -I$, we have $q=2$, so that 
  the radius of convergence of the series $Z(\varepsilon)$ is precisely
  \begin{equation}   \label{cex1}
       r = |\varepsilon_0^{(2)}| = \frac{2}{\alpha}.
\end{equation}
This result should be compared with the bound provided by the Magnus expansion. 
Since $\|A\| = \|B\| = 1$, Theorem \ref{conv-cbh} guarantees the
convergence of the BCH series in this case whenever $2 \alpha |\varepsilon| < \pi$, or
$|\varepsilon| < \frac{\pi}{2 \alpha}$, which, in view of (\ref{cex1}), is clearly a 
conservative estimate.

We can also check numerically the rate of convergence of the BCH series in this example as a function
of the parameter $\varepsilon$. Let us denote by $Z^{[N]}$ the sum of the first $N$ terms of
the series, i.e., 
\[
     Z^{[N]}(\varepsilon) = \sum_{n=1}^N Z_n(\varepsilon)
\]
and compute, for $\alpha = 2$ and different values of $\varepsilon$, the matrix
\[
    E_r(\varepsilon) = U(\varepsilon) \e^{-Z^{[N]}(\varepsilon)}  - I,
\]
where $U(\varepsilon)$ is given by (\ref{ex2.1}). If $\varepsilon$ belongs to the convergence
domain of the BCH series for the matrices $X$ and $Y$ (i.e., $|\varepsilon| < 1$), 
then $E_r(\varepsilon) \rightarrow 0$ as $N \rightarrow \infty$. 

First we take $\varepsilon = \frac{1}{4}$. With $N=10$, the elements of $E_r$ are of order
$10^{-7}$, whereas adding five additional terms in the series, $N=15$, the elements of $E_r$
are approximately $10^{-10}$.

Next we choose $\varepsilon = 0.9$, i.e., a value near the boundary of the convergence domain.
In this case with $N=15$ the convergence of the series does not manifest at all. In fact,
a much larger number of terms is required to achieve significant results. Thus, for the elements
of $E_r$ to be of order $10^{-8}$ we need to compute $N=150$ terms of the BCH series, whereas
with $N=200$ the elements of  $E_r$ are of order $10^{-10}$. The computations have been carried out
with the recurrence (\ref{varada}). $\Box$

As this example clearly shows, it is not always possible to determine accurately
the convergence domain of the BCH series by computing successive approximations,
since the rate of convergence can be slow indeed near the boundary. For this reason
it could be of interest to design a procedure to apply in practice the characterization of
the convergence in terms of the eigenvalues of the matrix $U(\varepsilon)$ analyzed  in
subsection \ref{cdeeig} for matrices.

This procedure could be as follows. Given two matrices $X$, $Y$, take the product of
exponentials
\[
      U(\varepsilon) = \e^{\varepsilon X} \, \e^{\varepsilon Y}
\]
with $\varepsilon = r \e^{i \theta}$. Next, define a grid in the $\varepsilon$-plane,
for instance in polar coordinates $(r,\theta)$, by $\Delta r = r_f /(n+1)$, $\Delta \theta = 2 \pi/(m+1)$
for two integers $n,m \ge 1$ and a sufficiently large value $r_f > 1$. Then, for each point
in the grid, $(r_k= k \Delta r, \theta_l = l \Delta \theta)$, $k=1,\ldots, n+1$, $l=0,1,\ldots, m$,
compute the corresponding matrix $U(\varepsilon)$ and its eigenstructure, locating where
there are multiple eigenvalues (within a  prescribed tolerance). If some of these multiple
eigenvalues have a negative real part, there exists a point in the neighborhood 
where the conditions enumerated in subsection \ref{cdeeig} are satisfied and therefore we
have approximately located the value of $\varepsilon$ where the BCH series fails to converge. 
This approximation can be made more accurate by applying, for instance, Newton's method.
The actual radius of convergence will be given by the smallest number $r$ found in this way.
Finally, if $r > 1$, then obviously the BCH series corresponding to $X$ and $Y$ converges.

\section{Some Applications}

As an illustration of the usefulness of the previous results, 
in this section we present two not so trivial applications of the formalism developed in section 2
for constructing explicitly the BCH series up to arbitrarily high order. 

\subsection{The symmetric BCH formula}

Sometimes it is necessary to compute the Lie series $W$ defined by 
\begin{equation}  \label{symbch.1}
    \exp(\frac{1}{2}X) \exp(Y) \exp(\frac{1}{2}X) = \exp(W).
\end{equation}
This occurs, for instance,  if one is interested in obtaining the order conditions
satisfied by time-symmetric composition methods for the numerical integration
of differential equations \cite{yoshida90coh,sanz-serna94nhp}. Two applications of the
usual BCH formula gives then the expression of $W$ in the Hall basis of
$\mathcal{L}(X,Y)$. 

A more efficient procedure is obtained, however, 
by introducing a parameter $\tau$ in (\ref{symbch.1}) such that
\begin{equation}  \label{symbch.2}
   W(\tau) = \log( \e^{\tau X/2} \, \e^{Y} \, \e^{\tau X/2})   
\end{equation}
and deriving the differential equation satisfied by $W(\tau)$. From the derivative of the
exponential map, one gets
\begin{equation}  \label{symbch.3}
   \frac{dW}{d\tau} = X + \sum_{n=2}^{\infty} \frac{B_n}{n!} \ad_W^n X, \qquad
     W(0) = Y
\end{equation}     
whence it is possible to construct explicitly $W$ as the series $W(\tau) = \sum_{k=0}^{\infty}
 W_{k}(\tau)$, with 
\begin{eqnarray}  \label{symbch.4} 
   W_1(\tau) & = & X \tau + Y  \nonumber  \\
   W_2(\tau) & = & 0   \\
   W_l(\tau) & = & \sum_{j=2}^{l-1} \frac{B_j}{j!} \int_0^{\tau} (\ad_W^j X)_l \, ds, 
   \qquad\quad l \ge 3 \nonumber
\end{eqnarray}
where, in general, $W_{2m} = 0$ for $m \ge 1$.
By following a similar approach as with equation (\ref{varada}) in the usual BCH series in
section 2, the
recursion (\ref{symbch.4}) allows one to express $W$ in (\ref{symbch.1}) as
\begin{equation}   \label{symbch.44}
   W =    \sum_{i \ge 1} w_i E_i.
\end{equation}
The coefficients $w_i$ of this series  up to degree 9 in the classical Hall basis are collected in Table 
\ref{tab:BCHsim}
in the Appendix. As with the usual BCH series, the coefficients up to degree 19 in both
Hall and Lyndon bases can be found at  \texttt{www.gicas.uji.es/research/bch.html}.

With respect to the convergence of the series, Theorem (\ref{conv-cbh}) guarantees that $W$ is 
convergent at least when $\|X\| + \|Y\| < \pi$.

\subsection{The BCH formula and a problem of R.C. Thompson}
\label{thompson}

In a series of papers \cite{thompson86poa,newman87nvo,thompson88sco,thompson89cpf},
R.C. Thompson considered the problem of constructing a
representation of the BCH formula as
\begin{equation}   \label{thomp.1}
   \e^X \, \e^Y = \e^Z   \qquad \mbox{ with } \qquad Z = S X S^{-1} + T Y T^{-1},
\end{equation}   
for certain functions $S = S(X,Y)$ and $T = T(X,Y)$ depending on $X$ and $Y$. By
using analytic techniques related with the Kashiwara--Vergne method, Rouvi\`ere 
\cite{rouviere86ese} proved 
that a Lie series $\rho(X,Y)$ exists such that
\begin{equation}    \label{thomp.2}
    S = \e^{\rho(X,Y)},  \qquad\quad  T = \e^{\rho(-Y,-X)}
\end{equation}
and converges when $X$, $Y$ are replaced by normed elements near $0$, whereas
the representation (\ref{thomp.1}) is global when both $X$ and $Y$ are skew-Hermitian matrices
\cite{thompson86poa}.

Thompson himself developed a computational technique for constructing explicitly the series
$\rho(X,Y)$ up to terms of degree ten. Although his results were not published,
he pointed out that they furnished strong evidence of the convergence of the series
$\rho(X,Y)$ on the closed unit sphere in any norm for which
$\| [X,Y] \| \le \|X\| \, \|Y\|$ \cite{thompson89cpf}. 

With the aim of clarifying this issue and illustrating the techniques developed in section 2,
we proceed next to compute $\rho(X,Y)$. Since $\rho(X,Y) \in \overline{\L(X,Y)}$, i.e., 
is a Lie series, it can be written as 
\[
   \rho(X,Y) = \sum_{i \ge 1} \rho_i E_i,
\]
where the elements $E_i$ have been introduced in (\ref{eq:BCHEi}), and 
the goal is to determine the coefficients $\rho_i$. This can be accomplished as follows. 
From the well known formula $\e^U V \e^{-U} = \e^{\ad_U} V$, it is clear that
\begin{equation}   \label{thomp.3}
   Z = \e^{\ad_{\rho(X,Y)}} X + \e^{\ad_{\rho(-Y,-X)}} Y.
\end{equation}
Next we expand $\e^{\ad_{\rho(X,Y)}} X$  and $\e^{\ad_{\rho(-Y,-X)}} Y$ into infinite
series as a linear combination of the Hall basis in $\mathcal{L}(X,Y)$ and match the resulting
terms with the corresponding to the BCH series for $Z$. Then a recursive system of equations is
obtained for the coefficients $\rho_i$. 

It is in fact possible to get a closed expression for
$\rho(X,Y)$ up to terms $Y^2$ by taking into account the corresponding formula of $Z$
\cite{postnikov94lga}. Specifically, from 
\begin{equation}  \label{mody2}
   Z = X + \frac{\ad_X}{1- \e^{-\ad_X}} Y  \quad \mathrm{mod}\; Y^2,
\end{equation}
a simple calculation leads to
\[
    \rho(X,Y) = f(\ad_X) Y  \quad \mathrm{mod}\; Y^2
\]
with the function $f(z)$ given by
\begin{equation}    \label{thomp.4}
   f(z) =  \frac{\e^z}{1- \e^z} + \frac{1}{z} \, \e^{\frac{1}{z}} = 
       -\frac{1}{4} - \frac{5}{96} z + \frac{1}{384} z^2 + \frac{143}{92160} z^3 +
    \frac{1}{122880} z^4 + \cdots
\end{equation}
Working in the classical Hall basis, the complete expression up to degree four reads
\begin{eqnarray*}
  \rho(X,Y) & = &  -\frac{1}{4} Y + \frac{5}{96} [Y,X] + \frac{1}{384} [[Y,X],X] + \frac{11}{768}
    [[Y,X],Y]  \\
     & & - \frac{143}{92160} [[[Y,X],X],X] - \frac{283}{92160} [[[Y,X],X],Y] +
    \frac{11}{23040} [[[Y,X],Y],Y]
\end{eqnarray*}
i.e., the corresponding equations have a unique solution. This is not the case, however,
at degree 5, where a free parameter appears, which can be chosen to be $\rho_{10}$. Then
\[
  \rho_{12} = \frac{-137-184320 \rho_{10}}{184320}, \quad   
    \rho_{13} =   \frac{-511-737280 \rho_{10}}{737280}
\]
As a matter of fact, if higher degrees are considered, more and more free parameters
appear in the corresponding solution. Thus, at degree 7 there are two additional parameters
(for instance, $\rho_{26}$ and $\rho_{30}$), whereas at degree 8 $\rho_{50}$ and $\rho_{52}$ 
can be chosen as free parameters.
We conclude, therefore, that there are infinite solutions to the problem posed by Thompson
depending on an increasing number of free parameters. An interesting issue would be 
to determine the value of these parameters in order to render the whole series convergent
on a domain as large as possible.

\subsection{Distribution of coefficients in the Lyndon basis}
\label{newsubs}

As we previously mentioned, there are noteworthy differences in the results obtained when
the algorithm of section \ref{section2} is applied to the BCH series
in the classical P. Hall basis and the Lyndon basis, particularly with respect to
number of vanishing coefficients. In the basis
of P. Hall there are 1316 zero coefficients out of 111013 up to degree $m=20$, whereas in the
Lyndon basis the number of vanishing terms rises to 34253 (more than $30 \%$ of the total
number of coefficients).

More remarkably, one notices that the distribution of these vanishing coefficients in the Lyndon basis follows a very specific pattern. Before entering into the details, let us denote for simplicity
$\mathcal{L}_m \equiv \mathcal{L}(X,Y)_m$. We first remark that, for each $m\geq 2$, the Lyndon basis $\mathcal{B}_m$ of $\mathcal{L}_m$ is a disjoint union $\mathcal{B}_m = \mathcal{B}_{m,1} \cup \mathcal{B}_{m,2}$ with $\mathcal{B}_{m,2} = [X,\mathcal{B}_{m-1}]$. Thus, $\mathcal{L}_m = \mathcal{L}_{m,1} \oplus \mathcal{L}_{m,2}$,
where $\mathcal{L}_{m,2} = [X, \mathcal{L}_{m-1}]$, and $\mathcal{B}_{m,k}$ ($k=1,2$) is a basis of $\mathcal{L}_{m,k}$. 
In particular, $\ad_X^{m-1} Y \in \mathcal{B}_m$.  In this sense, from our computations we make two observations: First,  the coefficient in the BCH formula of the element $\ad_X^{m-1} Y$ in the basis $\mathcal{B}_m$ is $0$ for even $m$. Second, the coefficients for the terms in $\mathcal{B}_{m,1}$ are also zero for even $m$. This gives a total number of
\[
   n_c(2p) = \dim(\mathcal{L}_{2p}) - \dim(\mathcal{L}_{2p-1}) + 1, \qquad p \ge 2
\]
vanishing coefficients of terms of degree $m=2p$ in the BCH formula written in the Lyndon basis. Thus, for instance, when $p=10$, the number of total number of vanishing coefficients is $n_c(20) =
\dim(\mathcal{L}_{20}) - \dim(\mathcal{L}_{19})+1 = 52377 - 27594+1 = 24784$.

With these considerations in mind, we can proceed next to explain the observed phenomena. 
 First, notice that expression (\ref{mody2}) gives explicitly the last term
of the BCH series in the Lyndon basis at each degree. By formally expanding in power series of
$\ad_X$ we get
\[
  Z = X + Y + \frac{1}{2} \ad_X Y + \sum_{k=2}^{\infty} \ad_X^k Y \quad \mathrm{mod}\; Y^2.
\]
Since $B_{2n+1} = 0$ for all $n \ge 1$,  the coefficient of $\ad_X^{k} Y$ is non-vanishing only
for odd values of $k$, or equivalently, for even values of the degree $m$. 

As for the remaining zero coefficients, let us consider at this point the symmetric BCH formula
(\ref{symbch.1}) again. Clearly the series (\ref{symbch.44}) only contains terms of odd
degree, i.e., $W = \sum_{i \ge 0} W_{2i+1}$, where $W_i \in  \mathcal{L}_i$.  By denoting
$P = X/2$ and forming the composition $\exp(P) \exp(W) \exp(-P)$ one gets trivially
\[
  \e^P \, \e^W \, \e^{-P} = \e^X \, \e^Y = \e^Z,
\]
i.e., the standard BCH formula. In the terminology of dynamical systems, $\exp(W)$ and
$\exp(Z)$ are said to be conjugated. Alternatively, we can write
$\exp(Z) = \exp(\ad_P) \, \exp(W)$, 
so that $Z = \exp(\ad_P) W$.  It is worth to write explicitly this relation for each term $Z_m \in
\mathcal{L}_m$ of the series $Z = \sum_{m \ge 0} Z_m$ by separating the odd and even degree
cases. Specifically, 
\begin{eqnarray*}
  Z_{2p+1} & = & W_{2p+1} + \sum_{j=1}^p \frac{1}{(2j)!) 2^{2j}} \, \ad_X^{2j} \, W_{2p-2j+1} \\
  Z_{2p} & = &  \sum_{j=1}^p \frac{1}{(2j-1)!) 2^{2j-1}} \, \ad_X^{2j-1} \, W_{2p-2j+1}. 
\end{eqnarray*}
From these expressions, it is clear that $Z_{2p+1}$ contains terms in the whole
subspace $\mathcal{L}_{2p+1,1} \oplus \mathcal{L}_{2p+1,2}$ (due to the presence of
$W_{2p+1}$),  whereas $Z_{2p}$ belongs to the subspace $\mathcal{L}_{2p,2}$, whose dimension is
equal to $\dim(\mathcal{L}_{2p-1})$. In other words, the remaining 
$\dim(\mathcal{L}_{2p}) - \dim(\mathcal{L}_{2p-1})$ \emph{must} necessarily vanish.
In this sense, the Lyndon basis seems the natural choice to get systematically the BCH series 
with the minimum number of terms. Nevertheless, compared to the basis of P. Hall, more CPU time and memory is required to compute the BCH with our algorithm in the Lyndon basis. In particular, 
 1.5 GBytes are required to compute the BCH formula up to degree 20 in the Hall basis, 
 whereas  3.6 GBytes of memory are needed in the Lyndon basis.

\section{Concluding remarks}

The effective computation of the BCH series has a long history and is closely related with 
the more general problem of carrying out symbolic computations in free Lie algebras. In this
work we have presented a new algorithm which allows us to get  a closed expression of the
series $Z = \log(\e^X \e^Y)$ up to degree 20 in terms of an arbitrary Hall--Viennot 
basis of the free Lie algebra generated by $X$ and $Y$, $\mathcal{L}(X,Y)$, requiring reasonable
computational resources. As far as we know, no other results are available
up to this degree in terms of a basis of $\mathcal{L}(X,Y)$. The algorithm is based on 
some more general results presented in \cite{murua06tha} on the connection of labeled 
rooted trees with an arbitrary Hall--Viennot basis of the free Lie algebra.

We have carried out explicitly the computations to get the coefficients of the BCH series in
terms of both the classical P. Hall basis and the Lyndon basis, with some noteworthy differences
in the corresponding results, as analyzed in subsection \ref{newsubs}.

We have also addressed the problem of the convergence of the series when $X$ and $Y$
are replaced by normed elements. In the particular case of $X$ and $Y$ being matrices, we
have provided a characterization of the convergence in terms of the eigenvalues of $\e^Z$.

Although here we have considered only the BCH series, it is clear that other more involved
calculations can be done, as is illustrated for instance by the problem of R.C. Thompson studied
in section \ref{thompson}. As a matter of fact, we intend to develop a general purpose package to carry
out symbolic computations in a free Lie algebra generated by more than two operators.

\subsection*{Acknowledgments}

The authors would like to thank Prof. Xavier Viennot for their very illuminating comments on the observed
pattern of zero coefficients in the Lyndon basis.
This work has been partially supported by Ministerio de
Educaci\'on y Ciencia (Spain) under project MTM2007-61572
(co-financed by the ERDF of the European Union) and Fundaci\'o Bancaixa. The SGI/IZO-SGIker UPV/EHU
(supported by the National Program for the Promotion of Human Resources within the National
Plan of Scientific Research, Development and Innovation - Fondo Social Europeo, MCyT and Basque
Government) is also gratefully acknowledged for generous allocation of resources for our computations
in the Lyndon basis.

\appendix

\section{Appendix}

In Table \ref{tab:BCH} we collect the indices $i'$ and $i'$ for $i\geq 3$ in (\ref{eq:Ei}) for the classical Hall basis and the values of the coefficients $z_i$ in the BCH formula (\ref{eq:BCHEi}) up to degree
9, whereas in Table \ref{tab:BCHsim} we gather the corresponding coefficients for the symmetric
BCH formula (\ref{symbch.1}).

\begin{table}
\caption{
Table of values of $i'$ and $i'$ for $i\geq 3$ in (\ref{eq:Ei}) for the classical Hall basis and the values $z_i \in \Q$ in the BCH formula (\ref{eq:BCHEi}). 
\label{tab:BCH}}
\begin{tabular}{|r|r|r|c||r|r|r|c||r|r|r|c|} \hline 
$i$ & $i'$ & $i''$ & $z_i$ &
$i$ & $i'$ & $i''$ & $z_i$ &
$i$ & $i'$ & $i''$ & $z_i$ \\ \hline
 1 &  &  & 1 & 44 & 25 & 2 &
   \ffrac{1}{10080} & 87 & 31 & 3 &
   -\ffrac{11}{30240} \\
 2 &  &  & 1 & 45 & 26 & 2 &
   \ffrac{23}{120960} & 88 & 32 & 3 &
   -\ffrac{19}{100800} \\ \cline{1-4}
 3 & 2 & 1 & -\ffrac{1}{2} & 46 & 27
   & 2 & \ffrac{1}{10080} & 89 & 33 &
   3 & -\ffrac{1}{43200} \\ \cline{1-4}
 4 & 3 & 1 & \ffrac{1}{12} & 47 & 28
   & 2 & \ffrac{1}{60480} & 90 & 34 &
   3 & -\ffrac{1}{10080} \\
 5 & 3 & 2 & -\ffrac{1}{12} & 48 & 29
   & 2 & 0 & 91 & 35 & 3 &
   -\ffrac{1}{50400} \\ \cline{1-4}
 6 & 4 & 1 & 0 & 49 & 15 & 3 & 0 &
   92 & 15 & 4 & -\ffrac{1}{33600} \\
 7 & 4 & 2 & \ffrac{1}{24} & 50 & 16
   & 3 & \ffrac{1}{4032} & 93 & 16 &
   4 & -\ffrac{13}{120960} \\
 8 & 5 & 2 & 0 & 51 & 17 & 3 &
   \ffrac{23}{30240} & 94 & 17 & 4 &
   -\ffrac{1}{10080} \\ \cline{1-4}
 9 & 6 & 1 & -\ffrac{1}{720} & 52 &
   18 & 3 & \ffrac{1}{2240} & 95 & 18
   & 4 & -\ffrac{11}{201600} \\
 10 & 6 & 2 & -\ffrac{1}{180} & 53 &
   19 & 3 & \ffrac{1}{15120} & 96 &
   19 & 4 & -\ffrac{1}{43200} \\
 11 & 7 & 2 & \ffrac{1}{180} & 54 &
   20 & 3 & 0 & 97 & 20 & 4 &
   -\ffrac{1}{7560} \\
 12 & 8 & 2 & \ffrac{1}{720} & 55 &
   21 & 3 & \ffrac{1}{2520} & 98 & 21
   & 4 & -\ffrac{1}{10080} \\
 13 & 4 & 3 & -\ffrac{1}{120} & 56 &
   22 & 3 & \ffrac{1}{10080} & 99 &
   22 & 4 & \ffrac{1}{50400} \\
 14 & 5 & 3 & -\ffrac{1}{360} & 57 &
   9 & 4 & 0 & 100 & 23 & 4 &
   \ffrac{1}{20160} \\ \cline{1-4}
 15 & 9 & 1 & 0 & 58 & 10 & 4 &
   \ffrac{1}{10080} & 101 & 15 & 5 &
   -\ffrac{23}{302400} \\
 16 & 9 & 2 & -\ffrac{1}{1440} & 59 &
   11 & 4 & -\ffrac{1}{20160} & 102 &
   16 & 5 & -\ffrac{1}{5760} \\
 17 & 10 & 2 & -\ffrac{1}{360} & 60 &
   12 & 4 & -\ffrac{1}{20160} & 103 &
   17 & 5 & \ffrac{13}{151200} \\
 18 & 11 & 2 & -\ffrac{1}{1440} & 61
   & 13 & 4 & 0 & 104 & 18 & 5 &
   \ffrac{19}{120960} \\
 19 & 12 & 2 & 0 & 62 & 14 & 4 &
   -\ffrac{1}{2520} & 105 & 19 & 5 &
   \ffrac{1}{33600} \\
 20 & 6 & 3 & 0 & 63 & 9 & 5 &
   \ffrac{1}{4032} & 106 & 20 & 5 &
   -\ffrac{13}{30240} \\
 21 & 7 & 3 & -\ffrac{1}{240} & 64 &
   10 & 5 & \ffrac{1}{840} & 107 & 21
   & 5 & -\ffrac{23}{100800} \\
 22 & 8 & 3 & -\ffrac{1}{720} & 65 &
   11 & 5 & \ffrac{1}{1440} & 108 &
   22 & 5 & -\ffrac{1}{100800} \\
 23 & 5 & 4 & \ffrac{1}{240} & 66 &
   12 & 5 & \ffrac{1}{12096} & 109 &
   23 & 5 & -\ffrac{1}{33600} \\ \cline{1-4}
 24 & 15 & 1 & \ffrac{1}{30240} & 67
   & 13 & 5 & \ffrac{1}{1260} & 110 &
   9 & 6 & -\ffrac{1}{60480} \\
 25 & 15 & 2 & \ffrac{1}{5040} & 68 &
   14 & 5 & \ffrac{1}{10080} & 111 &
   10 & 6 & -\ffrac{1}{90720} \\
 26 & 16 & 2 & \ffrac{1}{3780} & 69 &
   7 & 6 & -\ffrac{1}{10080} & 112 &
   11 & 6 & \ffrac{1}{30240} \\
 27 & 17 & 2 & -\ffrac{1}{3780} & 70
   & 8 & 6 & -\ffrac{13}{30240} & 113
   & 12 & 6 & -\ffrac{11}{302400} \\
 28 & 18 & 2 & -\ffrac{1}{5040} & 71
   & 8 & 7 & -\ffrac{1}{3360} & 114 &
   13 & 6 & \ffrac{1}{15120} \\ \cline{5-8}
 29 & 19 & 2 & -\ffrac{1}{30240} & 72
   & 42 & 1 & -\ffrac{1}{1209600} &
   115 & 14 & 6 & \ffrac{1}{3780} \\
 30 & 9 & 3 & \ffrac{1}{2016} & 73 &
   42 & 2 & -\ffrac{1}{151200} & 116
   & 9 & 7 & -\ffrac{11}{120960} \\
 31 & 10 & 3 & \ffrac{23}{15120} & 74
   & 43 & 2 & -\ffrac{1}{56700} & 117
   & 10 & 7 & -\ffrac{1}{6720} \\
 32 & 11 & 3 & \ffrac{1}{5040} & 75 &
   44 & 2 & -\ffrac{1}{75600} & 118 &
   11 & 7 & -\ffrac{1}{14400} \\
 33 & 12 & 3 & -\ffrac{1}{10080} & 76
   & 45 & 2 & \ffrac{1}{75600} & 119
   & 12 & 7 & -\ffrac{11}{120960} \\
 34 & 13 & 3 & \ffrac{1}{1260} & 77 &
   46 & 2 & \ffrac{1}{56700} & 120 &
   13 & 7 & -\ffrac{1}{20160} \\
 35 & 14 & 3 & \ffrac{1}{5040} & 78 &
   47 & 2 & \ffrac{1}{151200} & 121 &
   14 & 7 & \ffrac{17}{100800} \\
 36 & 6 & 4 & \ffrac{1}{5040} & 79 &
   48 & 2 & \ffrac{1}{1209600} & 122
   & 9 & 8 & -\ffrac{1}{20160} \\
 37 & 7 & 4 & -\ffrac{1}{10080} & 80
   & 24 & 3 & -\ffrac{1}{43200} & 123
   & 10 & 8 & \ffrac{17}{151200} \\
 38 & 8 & 4 & \ffrac{1}{1680} & 81 &
   25 & 3 & -\ffrac{37}{302400} & 124
   & 11 & 8 & \ffrac{1}{6048} \\
 39 & 6 & 5 & \ffrac{13}{15120} & 82
   & 26 & 3 & -\ffrac{11}{60480} &
   125 & 12 & 8 & \ffrac{1}{60480} \\
 40 & 7 & 5 & -\ffrac{1}{1120} & 83 &
   27 & 3 & -\ffrac{11}{302400} & 126
   & 13 & 8 & -\ffrac{1}{100800} \\
 41 & 8 & 5 & -\ffrac{1}{5040} & 84 &
   28 & 3 & \ffrac{11}{302400} & 127
   & 14 & 8 & \ffrac{1}{37800} \\ \cline{1-4}
 42 & 24 & 1 & 0 & 85 & 29 & 3 &
   \ffrac{1}{100800} &  & & \\
 43 & 24 & 2 & \ffrac{1}{60480} & 86
   & 30 & 3 & -\ffrac {1}{7560} &  & & \\ \hline 
\end{tabular} 

\end{table}

\begin{table}
\caption{
Table of values of $i'$ and $i'$ for $i\geq 3$ in (\ref{eq:Ei}) for the classical Hall basis and the values $w_i \in \Q$ in the symmetric BCH formula (\ref{symbch.1}).
\label{tab:BCHsim}}
\begin{tabular}{|r|r|r|c||r|r|r|c||r|r|r|c|} \hline
$i$ & $i'$ & $i''$ & $w_i$ &
$i$ & $i'$ & $i''$ & $w_i$ &
$i$ & $i'$ & $i''$ & $w_i$ \\ \hline
1 & 1 & 0 & 1 & 44 & 25 & 2 & 0 &
  87 & 31 & 3 & \ffrac{1}{4608} \\
2 & 2 & 0 & 1 & 45 & 26 & 2 & 0 &
  88 & 32 & 3 & \ffrac{23}{134400}
  \\ \cline{1-4}
3 & 2 & 1 & 0 & 46 & 27 & 2 & 0 &
  89 & 33 & 3 & \ffrac{1}{37800} \\ \cline{1-4}
4 & 3 & 1 & -\ffrac{1}{24} & 47 & 28
  & 2 & 0 & 90 & 34 & 3 &
  \ffrac{1}{23040} \\
5 & 3 & 2 & -\ffrac{1}{12} & 48 & 29
  & 2 & 0 & 91 & 35 & 3 &
  \ffrac{1}{201600} \\ \cline{1-4}
6 & 4 & 1 & 0 & 49 & 15 & 3 & 0 &
  92 & 15 & 4 & \ffrac{193}{6451200}
  \\
7 & 4 & 2 & 0 & 50 & 16 & 3 & 0 &
  93 & 16 & 4 & \ffrac{53}{483840}
  \\
8 & 5 & 2 & 0 & 51 & 17 & 3 & 0 &
  94 & 17 & 4 & \ffrac{25}{193536}
  \\ \cline{1-4}
9 & 6 & 1 & \ffrac{7}{5760} & 52 &
  18 & 3 & 0 & 95 & 18 & 4 &
  \ffrac{1}{22400} \\
10 & 6 & 2 & \ffrac{7}{1440} & 53 &
  19 & 3 & 0 & 96 & 19 & 4 &
  -\ffrac{13}{1209600} \\
11 & 7 & 2 & \ffrac{1}{180} & 54 &
  20 & 3 & 0 & 97 & 20 & 4 &
  \ffrac{53}{483840} \\
12 & 8 & 2 & \ffrac{1}{720} & 55 &
  21 & 3 & 0 & 98 & 21 & 4 &
  \ffrac{17}{161280} \\
13 & 4 & 3 & \ffrac{1}{480} & 56 &
  22 & 3 & 0 & 99 & 22 & 4 &
  -\ffrac{3}{44800} \\
14 & 5 & 3 & -\ffrac{1}{360} & 57 &
  9 & 4 & 0 & 100 & 23 & 4 &
  -\ffrac{19}{322560} \\ \cline{1-4}
15 & 9 & 1 & 0 & 58 & 10 & 4 & 0 &
  101 & 15 & 5 &
  \ffrac{367}{4838400} \\
16 & 9 & 2 & 0 & 59 & 11 & 4 & 0 &
  102 & 16 & 5 & \ffrac{193}{645120}
  \\
17 & 10 & 2 & 0 & 60 & 12 & 4 & 0 &
  103 & 17 & 5 & \ffrac{247}{604800}
  \\
18 & 11 & 2 & 0 & 61 & 13 & 4 & 0 &
  104 & 18 & 5 & \ffrac{53}{241920}
  \\
19 & 12 & 2 & 0 & 62 & 14 & 4 & 0 &
  105 & 19 & 5 & \ffrac{1}{33600} \\
20 & 6 & 3 & 0 & 63 & 9 & 5 & 0 &
  106 & 20 & 5 & \ffrac{53}{161280}
  \\
21 & 7 & 3 & 0 & 64 & 10 & 5 & 0 &
  107 & 21 & 5 & \ffrac{193}{403200}
  \\
22 & 8 & 3 & 0 & 65 & 11 & 5 & 0 &
  108 & 22 & 5 & \ffrac{13}{201600}
  \\
23 & 5 & 4 & 0 & 66 & 12 & 5 & 0 &
  109 & 23 & 5 & -\ffrac{1}{5600} \\ \cline{1-4}
24 & 15 & 1 & -\ffrac{31}{967680} &
  67 & 13 & 5 & 0 & 110 & 9 & 6 &
  \ffrac{11}{774144} \\
25 & 15 & 2 & -\ffrac{31}{161280} &
  68 & 14 & 5 & 0 & 111 & 10 & 6 &
  \ffrac{1}{290304} \\
26 & 16 & 2 & -\ffrac{13}{30240} &
  69 & 7 & 6 & 0 & 112 & 11 & 6 &
  -\ffrac{1}{15360} \\
27 & 17 & 2 & -\ffrac{53}{120960} &
  70 & 8 & 6 & 0 & 113 & 12 & 6 &
  -\ffrac{89}{1209600} \\
28 & 18 & 2 & -\ffrac{1}{5040} & 71
  & 8 & 7 & 0 & 114 & 13 & 6 &
  -\ffrac{11}{241920} \\ \cline{5-8}
29 & 19 & 2 & -\ffrac{1}{30240} & 72
  & 42 & 1 & \ffrac{127}{154828800}
  & 115 & 14 & 6 &
  -\ffrac{13}{80640} \\
30 & 9 & 3 & -\ffrac{53}{161280} &
  73 & 42 & 2 &
  \ffrac{127}{19353600} & 116 & 9 &
  7 & \ffrac{1}{12096} \\
31 & 10 & 3 & -\ffrac{11}{12096} &
  74 & 43 & 2 & \ffrac{157}{7257600}
  & 117 & 10 & 7 & \ffrac{11}{64512}
  \\
32 & 11 & 3 & -\ffrac{3}{4480} & 75
  & 44 & 2 & \ffrac{367}{9676800} &
  118 & 11 & 7 & \ffrac{1}{33600} \\
33 & 12 & 3 & -\ffrac{1}{10080} & 76
  & 45 & 2 & \ffrac{23}{604800} &
  119 & 12 & 7 & -\ffrac{11}{120960}
  \\
34 & 13 & 3 & -\ffrac{1}{4032} & 77
  & 46 & 2 & \ffrac{79}{3628800} &
  120 & 13 & 7 & \ffrac{1}{35840} \\
35 & 14 & 3 & -\ffrac{1}{6720} & 78
  & 47 & 2 & \ffrac{1}{151200} & 121
  & 14 & 7 & -\ffrac{29}{134400} \\
36 & 6 & 4 & -\ffrac{19}{80640} & 79
  & 48 & 2 & \ffrac{1}{1209600} &
  122 & 9 & 8 & \ffrac{211}{1935360}
  \\
37 & 7 & 4 & -\ffrac{1}{10080} & 80
  & 24 & 3 & \ffrac{367}{19353600} &
  123 & 10 & 8 & \ffrac{173}{604800}
  \\
38 & 8 & 4 & \ffrac{17}{40320} & 81
  & 25 & 3 & \ffrac{473}{4838400} &
  124 & 11 & 8 & \ffrac{5}{24192} \\
39 & 6 & 5 & -\ffrac{53}{60480} & 82
  & 26 & 3 & \ffrac{41}{215040} &
  125 & 12 & 8 & \ffrac{1}{60480} \\
40 & 7 & 5 & -\ffrac{19}{13440} & 83
  & 27 & 3 & \ffrac{211}{1209600} &
  126 & 13 & 8 & \ffrac{61}{403200}
  \\
41 & 8 & 5 & -\ffrac{1}{5040} & 84 &
  28 & 3 & \ffrac{89}{1209600} & 127
  & 14 & 8 & -\ffrac{1}{151200} \\ \cline{1-4}
42 & 24 & 1 & 0 & 85 & 29 & 3 &
  \ffrac{1}{100800} &  &  & \\
43 & 24 & 2 & 0 & 86 & 30 & 3 &
  \ffrac{79}{967680} &  &  & \\ \hline
\end{tabular}

\end{table}

\newpage

\bibliographystyle{plain}
\bibliography{ourbib,geom_int,numerbib,nirea}

\end{document}

%% file: 2trees.tex
 \newcommand{\nta}[1][1.5ex]{%
  \mbox{\setlength{\unitlength}{#1}%
  \begin{picture}(0.63,0.63)(-0.315,-0.315)
  \put (0,0){\circle*{0.6}}
  \end{picture}}}

  \newcommand{\ntb}[1][1.5ex]{%
  \mbox{\setlength{\unitlength}{#1}%
  \begin{picture}(0.63,0.63)(-0.315,-0.315)
  \put (0,0){\circle{0.6}}
  \end{picture}}}

  \newcommand{\ntc}[1][1.5ex]{%
  \mbox{\setlength{\unitlength}{#1}%
  \begin{picture}(0.63,1.68)(-0.315,-0.34)
  \put (0,0.3){\line(0,1){0.4}}
  \put (0,0){\circle*{0.6}}
  \put (0,1){\circle*{0.6}}
  \end{picture}}}

  \newcommand{\ntd}[1][1.5ex]{%
  \mbox{\setlength{\unitlength}{#1}%
  \begin{picture}(0.63,1.68)(-0.315,-0.34)
  \put (0,0.3){\line(0,1){0.4}}
  \put (0,0){\circle*{0.6}}
  \put (0,1){\circle{0.6}}
  \end{picture}}}

  \newcommand{\nte}[1][1.5ex]{%
  \mbox{\setlength{\unitlength}{#1}%
  \begin{picture}(0.63,1.68)(-0.315,-0.34)
  \put (0,0.3){\line(0,1){0.4}}
  \put (0,0){\circle{0.6}}
  \put (0,1){\circle*{0.6}}
  \end{picture}}}

  \newcommand{\ntf}[1][1.5ex]{%
  \mbox{\setlength{\unitlength}{#1}%
  \begin{picture}(0.63,1.68)(-0.315,-0.34)
  \put (0,0.3){\line(0,1){0.4}}
  \put (0,0){\circle{0.6}}
  \put (0,1){\circle{0.6}}
  \end{picture}}}

  \newcommand{\ntg}[1][1.5ex]{%
  \mbox{\setlength{\unitlength}{#1}%
  \begin{picture}(1.575,2.73)(-1.2375,-0.365)
  \put (-0.200689,0.222988){\line(-5,6){0.498621}}
  \put (-0.699311,1.22299){\line(5,6){0.498621}}
  \put (0,0){\circle*{0.6}}
  \put (-0.9,1){\circle*{0.6}}
  \put (0,2){\circle*{0.6}}
  \end{picture}}}

  \newcommand{\nth}[1][1.5ex]{%
  \mbox{\setlength{\unitlength}{#1}%
  \begin{picture}(1.575,2.73)(-1.2375,-0.365)
  \put (-0.200689,0.222988){\line(-5,6){0.498621}}
  \put (-0.699311,1.22299){\line(5,6){0.498621}}
  \put (0,0){\circle*{0.6}}
  \put (-0.9,1){\circle*{0.6}}
  \put (0,2){\circle{0.6}}
  \end{picture}}}

  \newcommand{\nti}[1][1.5ex]{%
  \mbox{\setlength{\unitlength}{#1}%
  \begin{picture}(1.575,2.73)(-1.2375,-0.365)
  \put (-0.200689,0.222988){\line(-5,6){0.498621}}
  \put (-0.699311,1.22299){\line(5,6){0.498621}}
  \put (0,0){\circle*{0.6}}
  \put (-0.9,1){\circle{0.6}}
  \put (0,2){\circle*{0.6}}
  \end{picture}}}

  \newcommand{\ntao}[1][1.5ex]{%
  \mbox{\setlength{\unitlength}{#1}%
  \begin{picture}(1.575,2.73)(-1.2375,-0.365)
  \put (-0.200689,0.222988){\line(-5,6){0.498621}}
  \put (-0.699311,1.22299){\line(5,6){0.498621}}
  \put (0,0){\circle*{0.6}}
  \put (-0.9,1){\circle{0.6}}
  \put (0,2){\circle{0.6}}
  \end{picture}}}

  \newcommand{\ntac}[1][1.5ex]{%
  \mbox{\setlength{\unitlength}{#1}%
  \begin{picture}(1.575,2.73)(-1.2375,-0.365)
  \put (-0.200689,0.222988){\line(-5,6){0.498621}}
  \put (-0.699311,1.22299){\line(5,6){0.498621}}
  \put (0,0){\circle{0.6}}
  \put (-0.9,1){\circle{0.6}}
  \put (0,2){\circle*{0.6}}
  \end{picture}}}

  \newcommand{\ntae}[1][1.5ex]{%
  \mbox{\setlength{\unitlength}{#1}%
  \begin{picture}(2.52,1.68)(-1.26,-0.34)
  \put (-0.200689,0.222988){\line(-5,6){0.498621}}
  \put (0.200689,0.222988){\line(5,6){0.498621}}
  \put (-0.9,1){\circle*{0.6}}
  \put (0.9,1){\circle*{0.6}}
  \put (0,0){\circle*{0.6}}
  \end{picture}}}

  \newcommand{\ntaf}[1][1.5ex]{%
  \mbox{\setlength{\unitlength}{#1}%
  \begin{picture}(2.52,1.68)(-1.26,-0.34)
  \put (-0.200689,0.222988){\line(-5,6){0.498621}}
  \put (0.200689,0.222988){\line(5,6){0.498621}}
  \put (-0.9,1){\circle*{0.6}}
  \put (0.9,1){\circle{0.6}}
  \put (0,0){\circle*{0.6}}
  \end{picture}}}

  \newcommand{\ntag}[1][1.5ex]{%
  \mbox{\setlength{\unitlength}{#1}%
  \begin{picture}(2.52,1.68)(-1.26,-0.34)
  \put (-0.200689,0.222988){\line(-5,6){0.498621}}
  \put (0.200689,0.222988){\line(5,6){0.498621}}
  \put (-0.9,1){\circle{0.6}}
  \put (0.9,1){\circle{0.6}}
  \put (0,0){\circle*{0.6}}
  \end{picture}}}

  \newcommand{\ntah}[1][1.5ex]{%
  \mbox{\setlength{\unitlength}{#1}%
  \begin{picture}(2.52,1.68)(-1.26,-0.34)
  \put (-0.200689,0.222988){\line(-5,6){0.498621}}
  \put (0.200689,0.222988){\line(5,6){0.498621}}
  \put (-0.9,1){\circle*{0.6}}
  \put (0.9,1){\circle*{0.6}}
  \put (0,0){\circle{0.6}}
  \end{picture}}}

  \newcommand{\ntai}[1][1.5ex]{%
  \mbox{\setlength{\unitlength}{#1}%
  \begin{picture}(2.52,1.68)(-1.26,-0.34)
  \put (-0.200689,0.222988){\line(-5,6){0.498621}}
  \put (0.200689,0.222988){\line(5,6){0.498621}}
  \put (-0.9,1){\circle*{0.6}}
  \put (0.9,1){\circle{0.6}}
  \put (0,0){\circle{0.6}}
  \end{picture}}}

\newcommand{\ntbb}[1][1.5ex]{%
\mbox{\setlength{\unitlength}{#1}%
\begin{picture}(1.7325,3.78)(-1.39125,-0.39)
\put (-0.217241,0.206897){\line(-1,1){0.615517}}
\put (-0.832759,1.2069){\line(1,1){0.615517}}
\put (0,0){\circle*{0.6}}
\put (-1.05,1){\circle*{0.6}}
\put (-0.200258,2.22338){\line(-5,6){0.49599}}
\put (0,2){\circle*{0.6}}
\put (-0.896506,3){\circle{0.6}}
\end{picture}}}

\newcommand{\ntca}[1][1.5ex]{%
\mbox{\setlength{\unitlength}{#1}%
\begin{picture}(2.72475,2.73)(-2.35988,-0.365)
\put (-0.217241,0.206897){\line(-1,1){0.615517}}
\put (0,0){\circle*{0.6}}
\put (-1.25605,1.21804){\line(-1,1){0.532898}}
\put (-0.843949,1.21804){\line(1,1){0.532898}}
\put (-1.995,2){\circle{0.6}}
\put (-0.105,2){\circle{0.6}}
\put (-1.05,1){\circle*{0.6}}
\end{picture}}}

  \newcommand{\ntdf}[1][1.5ex]{%
  \mbox{\setlength{\unitlength}{#1}%
  \begin{picture}(2.44913,2.73)(-2.09081,-0.365)
  \put (-0.217241,0.206897){\line(-1,1){0.615517}}
  \put (0,0){\circle{0.6}}
  \put (-1.21912,1.24779){\line(-2,3){0.344267}}
  \put (-0.880884,1.24779){\line(2,3){0.344267}}
  \put (-1.7325,2){\circle*{0.6}}
  \put (-0.3675,2){\circle*{0.6}}
  \put (-1.05,1){\circle{0.6}}
  \end{picture}}}

  \newcommand{\nted}[1][1.5ex]{%
  \mbox{\setlength{\unitlength}{#1}%
  \begin{picture}(3.15,2.73)(-1.575,-0.365)
  \put (-0.230466,0.192055){\line(-6,5){0.739067}}
  \put (0.230466,0.192055){\line(6,5){0.739067}}
  \put (-1.2,1){\circle{0.6}}
  \put (1.01549,1.23655){\line(-4,5){0.410981}}
  \put (1.2,1){\circle*{0.6}}
  \put (0.42,2){\circle{0.6}}
  \put (0,0){\circle*{0.6}}
  \end{picture}}}

  \newcommand{\ntei}[1][1.5ex]{%
  \mbox{\setlength{\unitlength}{#1}%
  \begin{picture}(3.15,2.73)(-1.575,-0.365)
  \put (-0.230466,0.192055){\line(-6,5){0.739067}}
  \put (0.230466,0.192055){\line(6,5){0.739067}}
  \put (-1.2,1){\circle*{0.6}}
  \put (1.01549,1.23655){\line(-4,5){0.410981}}
  \put (1.2,1){\circle{0.6}}
  \put (0.42,2){\circle*{0.6}}
  \put (0,0){\circle{0.6}}
  \end{picture}}}

  \newcommand{\ntfh}[1][1.5ex]{%
  \mbox{\setlength{\unitlength}{#1}%
  \begin{picture}(3.15,1.68)(-1.575,-0.34)
  \put (-0.230466,0.192055){\line(-6,5){0.739067}}
  \put (0.,0.3){\line(0,1){0.4}}
  \put (0.230466,0.192055){\line(6,5){0.739067}}
  \put (-1.2,1){\circle{0.6}}
  \put (0.,1){\circle{0.6}}
  \put (1.2,1){\circle{0.6}}
  \put (0,0){\circle*{0.6}}
  \end{picture}}}

  \newcommand{\ntfi}[1][1.5ex]{%
  \mbox{\setlength{\unitlength}{#1}%
  \begin{picture}(3.15,1.68)(-1.575,-0.34)
  \put (-0.230466,0.192055){\line(-6,5){0.739067}}
  \put (0.,0.3){\line(0,1){0.4}}
  \put (0.230466,0.192055){\line(6,5){0.739067}}
  \put (-1.2,1){\circle*{0.6}}
  \put (0.,1){\circle*{0.6}}
  \put (1.2,1){\circle*{0.6}}
  \put (0,0){\circle{0.6}}
  \end{picture}}}

  \newcommand{\ntgo}[1][1.5ex]{%
  \mbox{\setlength{\unitlength}{#1}%
  \begin{picture}(3.15,1.68)(-1.575,-0.34)
  \put (-0.230466,0.192055){\line(-6,5){0.739067}}
  \put (0.,0.3){\line(0,1){0.4}}
  \put (0.230466,0.192055){\line(6,5){0.739067}}
  \put (-1.2,1){\circle*{0.6}}
  \put (0.,1){\circle*{0.6}}
  \put (1.2,1){\circle{0.6}}
  \put (0,0){\circle{0.6}}
  \end{picture}}}

  \newcommand{\ntga}[1][1.5ex]{%
  \mbox{\setlength{\unitlength}{#1}%
  \begin{picture}(3.15,1.68)(-1.575,-0.34)
  \put (-0.230466,0.192055){\line(-6,5){0.739067}}
  \put (0.,0.3){\line(0,1){0.4}}
  \put (0.230466,0.192055){\line(6,5){0.739067}}
  \put (-1.2,1){\circle*{0.6}}
  \put (0.,1){\circle{0.6}}
  \put (1.2,1){\circle{0.6}}
  \put (0,0){\circle{0.6}}
  \end{picture}}}

\newcommand{\ntbfg}[1][1.5ex]{%
\mbox{\setlength{\unitlength}{#1}%
\begin{picture}(3.78,2.73)(-1.89,-0.365)
\put (-0.249615,0.16641){\line(-3,2){1.00077}}
\put (0.,0.3){\line(0,1){0.4}}
\put (0.249615,0.16641){\line(3,2){1.00077}}
\put (-1.5,1){\circle*{0.6}}
\put (0.,1){\circle*{0.6}}
\put (1.37689,1.27358){\line(-1,2){0.203781}}
\put (1.5,1){\circle{0.6}}
\put (1.05,2){\circle*{0.6}}
\put (0,0){\circle{0.6}}
\end{picture}}}

  \newcommand{\ntbga}[1][1.5ex]{%
  \mbox{\setlength{\unitlength}{#1}%
  \begin{picture}(3.78,2.73)(-1.89,-0.365)
  \put (-0.249615,0.16641){\line(-3,2){1.00077}}
  \put (0.,0.3){\line(0,1){0.4}}
  \put (0.249615,0.16641){\line(3,2){1.00077}}
  \put (-1.5,1){\circle*{0.6}}
  \put (0.,1){\circle{0.6}}
  \put (1.36854,1.26966){\line(-1,2){0.224579}}
  \put (1.5,1){\circle{0.6}}
  \put (1.0125,2){\circle*{0.6}}
  \put (0,0){\circle{0.6}}
  \end{picture}}}

  \newcommand{\ntbhb}[1][1.5ex]{%
  \mbox{\setlength{\unitlength}{#1}%
  \begin{picture}(3.78,1.68)(-1.89,-0.34)
  \put (-0.249615,0.16641){\line(-3,2){1.00077}}
  \put (-0.134164,0.268328){\line(-1,2){0.231672}}
  \put (0.134164,0.268328){\line(1,2){0.231672}}
  \put (0.249615,0.16641){\line(3,2){1.00077}}
  \put (-1.5,1){\circle*{0.6}}
  \put (-0.5,1){\circle*{0.6}}
  \put (0.5,1){\circle*{0.6}}
  \put (1.5,1){\circle*{0.6}}
  \put (0,0){\circle{0.6}}
  \end{picture}}}

  \newcommand{\ntbhc}[1][1.5ex]{%
  \mbox{\setlength{\unitlength}{#1}%
  \begin{picture}(3.78,1.68)(-1.89,-0.34)
  \put (-0.249615,0.16641){\line(-3,2){1.00077}}
  \put (-0.134164,0.268328){\line(-1,2){0.231672}}
  \put (0.134164,0.268328){\line(1,2){0.231672}}
  \put (0.249615,0.16641){\line(3,2){1.00077}}
  \put (-1.5,1){\circle*{0.6}}
  \put (-0.5,1){\circle*{0.6}}
  \put (0.5,1){\circle*{0.6}}
  \put (1.5,1){\circle{0.6}}
  \put (0,0){\circle{0.6}}
  \end{picture}}}

  \newcommand{\ntbhd}[1][1.5ex]{%
  \mbox{\setlength{\unitlength}{#1}%
  \begin{picture}(3.78,1.68)(-1.89,-0.34)
  \put (-0.249615,0.16641){\line(-3,2){1.00077}}
  \put (-0.134164,0.268328){\line(-1,2){0.231672}}
  \put (0.134164,0.268328){\line(1,2){0.231672}}
  \put (0.249615,0.16641){\line(3,2){1.00077}}
  \put (-1.5,1){\circle*{0.6}}
  \put (-0.5,1){\circle*{0.6}}
  \put (0.5,1){\circle{0.6}}
  \put (1.5,1){\circle{0.6}}
  \put (0,0){\circle{0.6}}
  \end{picture}}}

  \newcommand{\ntbhe}[1][1.5ex]{%
  \mbox{\setlength{\unitlength}{#1}%
  \begin{picture}(3.78,1.68)(-1.89,-0.34)
  \put (-0.249615,0.16641){\line(-3,2){1.00077}}
  \put (-0.134164,0.268328){\line(-1,2){0.231672}}
  \put (0.134164,0.268328){\line(1,2){0.231672}}
  \put (0.249615,0.16641){\line(3,2){1.00077}}
  \put (-1.5,1){\circle*{0.6}}
  \put (-0.5,1){\circle{0.6}}
  \put (0.5,1){\circle{0.6}}
  \put (1.5,1){\circle{0.6}}
  \put (0,0){\circle{0.6}}
  \end{picture}}}

\newcommand{\ntgd}[1][1.5ex]{%
\mbox{\setlength{\unitlength}{#1}%
\begin{picture}(1.81125,4.83)(-1.46813,-0.415)
\put (-0.224223,0.199309){\line(-6,5){0.676554}}
\put (-0.900777,1.19931){\line(6,5){0.676554}}
\put (0,0){\circle*{0.6}}
\put (-1.125,1){\circle*{0.6}}
\put (-0.207821,2.21636){\line(-1,1){0.544901}}
\put (-0.752721,3.21636){\line(1,1){0.544901}}
\put (0,2){\circle*{0.6}}
\put (-0.960542,3){\circle*{0.6}}
\put (0,4){\circle{0.6}}
\end{picture}}}
\newcommand{\nthc}[1][1.5ex]{%
\mbox{\setlength{\unitlength}{#1}%
\begin{picture}(2.81982,3.78)(-1.49214,-0.39)
\put (-0.224223,0.199309){\line(-6,5){0.676554}}
\put (-0.900777,1.19931){\line(6,5){0.676554}}
\put (0,0){\circle*{0.6}}
\put (-1.125,1){\circle*{0.6}}
\put (-0.207821,2.21636){\line(-1,1){0.544901}}
\put (0.207821,2.21636){\line(1,1){0.544901}}
\put (-0.960542,3){\circle{0.6}}
\put (0.960542,3){\circle{0.6}}
\put (0,2){\circle*{0.6}}
\end{picture}}}
\newcommand{\ntabo}[1][1.5ex]{%
\mbox{\setlength{\unitlength}{#1}%
\begin{picture}(3.024,2.73)(-2.652,-0.365)
\put (-0.230466,0.192055){\line(-6,5){0.739067}}
\put (0,0){\circle*{0.6}}
\put (-1.42013,1.20382){\line(-1,1){0.639744}}
\put (-1.2,1.3){\line(0,1){0.4}}
\put (-0.979872,1.20382){\line(1,1){0.639744}}
\put (-2.28,2){\circle{0.6}}
\put (-1.2,2){\circle{0.6}}
\put (-0.12,2){\circle{0.6}}
\put (-1.2,1){\circle*{0.6}}
\end{picture}}}
\newcommand{\ntboa}[1][1.5ex]{%
\mbox{\setlength{\unitlength}{#1}%
\begin{picture}(4.74075,2.73)(-1.76288,-0.365)
\put (-0.241067,0.178568){\line(-4,3){0.867866}}
\put (0.241067,0.178568){\line(4,3){0.867866}}
\put (-1.35,1){\circle{0.6}}
\put (1.11837,1.19065){\line(-6,5){0.751732}}
\put (1.58163,1.19065){\line(6,5){0.751732}}
\put (0.135,2){\circle{0.6}}
\put (2.565,2){\circle{0.6}}
\put (1.35,1){\circle*{0.6}}
\put (0,0){\circle*{0.6}}
\end{picture}}}
\newcommand{\ntbha}[1][1.5ex]{%
\mbox{\setlength{\unitlength}{#1}%
\begin{picture}(3.78,1.68)(-1.89,-0.34)
\put (-0.249615,0.16641){\line(-3,2){1.00077}}
\put (-0.134164,0.268328){\line(-1,2){0.231672}}
\put (0.134164,0.268328){\line(1,2){0.231672}}
\put (0.249615,0.16641){\line(3,2){1.00077}}
\put (-1.5,1){\circle{0.6}}
\put (-0.5,1){\circle{0.6}}
\put (0.5,1){\circle{0.6}}
\put (1.5,1){\circle{0.6}}
\put (0,0){\circle*{0.6}}
\end{picture}}}
\newcommand{\ntbcg}[1][1.5ex]{%
\mbox{\setlength{\unitlength}{#1}%
\begin{picture}(3.15,2.73)(-1.575,-0.365)
\put (-0.230466,0.192055){\line(-6,5){0.739067}}
\put (0.230466,0.192055){\line(6,5){0.739067}}
\put (-1.01549,1.23655){\line(4,5){0.410981}}
\put (-1.2,1){\circle*{0.6}}
\put (-0.42,2){\circle{0.6}}
\put (1.01549,1.23655){\line(-4,5){0.410981}}
\put (1.2,1){\circle*{0.6}}
\put (0.42,2){\circle{0.6}}
\put (0,0){\circle*{0.6}}
\end{picture}}}